\documentclass[12pt,notitlepage,aps,amsmath,amssymb]{revtex4-1}

\usepackage{bm}
\usepackage{mathtools}
\usepackage{subfig}
\usepackage{graphicx}
\usepackage{caption}

\usepackage{times}
\usepackage{newtxtext}

\newcommand{\Refs}[1]{Refs.~\onlinecite{#1}}
\newcommand{\Eqref}[1]{Eq.~\eqref{#1}}
\newcommand{\Eqsref}[1]{Eqs.~\eqref{#1}}
\newcommand{\Figref}[1]{Fig.~\ref{#1}}

\newcommand{\Secref}[1]{Sec.~\ref{#1}}
\newcommand{\Appref}[1]{App.~\ref{#1}}

\newcommand{\Sn}{\Phi} 
\newcommand{\sn}{\phi} 
\newcommand{\Snw}{\varphi} 

\newcommand{\Snn}{\Psi}

\newcommand{\nSnn}{{\widetilde{\Snn}}}

\newcommand{\nSn}{{\widetilde{\Sn}}}
\newcommand{\nsn}{{\widetilde{\sn}}}

\newcommand{\abs}[1]{\lvert {#1} \rvert}
\newcommand{\mean}[1]{\left< {#1} \right>}
\newcommand{\tmean}[1]{\langle {#1} \rangle} 

\newcommand{\mSn}{\left< \Sn \right>} 
\newcommand{\rSn}{\Sn_{\text{rms}}} 
\newcommand{\mSnn}{\left< \Snn \right>} 
\newcommand{\rSnn}{\Snn_{\text{rms}}} 

\newcommand{\rms}{\text{rms}}
\newcommand{\rmd}{\text{d}} 
\newcommand{\td}{\tau_{\text{d}}} 
\newcommand{\tw}{\tau_{\text{w}}} 

\begin{document}

\title{Probability distribution functions for intermittent scrape-off layer plasma fluctuations}

\author{A.~Theodorsen}
\email{audun.theodorsen@uit.no}
\affiliation{Department of Physics and Technology, UiT The Arctic University of Norway, N-9037 Troms{\o}, Norway}

\author{O.~E.~Garcia}
\email{odd.erik.garcia@uit.no}
\affiliation{Department of Physics and Technology, UiT The Arctic University of Norway, N-9037 Troms{\o}, Norway}

\date{\today}

\begin{abstract}
  A stochastic model for intermittent fluctuations in the scrape-off layer of magnetically confined plasmas has been constructed based on a super-position of uncorrelated pulses arriving according to a Poisson process. In the most common applications of the model, the pulse amplitudes are assumed exponentially distributed, supported by conditional averaging of large-amplitude fluctuations in experimental measurement data. This basic assumption has two potential limitations. First, statistical analysis of measurement data using conditional averaging only reveals the tail of the amplitude distribution to be exponentially distributed. Second, exponentially distributed amplitudes leads to a positive definite signal which cannot capture fluctuations in for example electric potential and radial velocity. Assuming pulse amplitudes which are not positive definite often make finding a closed form for the probability density function difficult, even if the characteristic function remains relatively simple. Thus estimating model parameters requires an approach based on the characteristic function, not the probability density function. In this contribution, the effect of changing the amplitude distribution on the moments, probability density function and characteristic function of the process is investigated and a parameter estimation method using the empirical distribution function is presented and tested on synthetically generated data. This proves valuable for describing intermittent fluctuations in the boundary region of magnetized plasmas.
\end{abstract}

\maketitle

\section{Introduction}

Radial propagation of filamentary structures is the main contributor to cross-field transport of particles and heat in the scrape-off layer (SOL) of magnetically confined plasmas \cite{labombard-2001,krasheninnikov-2001,rudakov-2002,dippolito-2004,rudakov-2005,zweben-2007,krasheninnikov-2008,garcia-2009,dippolito-2011,kube-2013,zweben-2015}. This turbulence-driven transport results in broad plasma profiles and enhanced plasma-wall interactions \cite{rudakov-2005,pitts-2005,lipschultz-2007,garcia-jnm-2007,garcia-nf-2007,militello-ppcf-2013,carralero-nf-2014,carralero-prl-2015,militello-ppcf-2016,walkden-ppcf-2017}.  

Statistical analysis of single-point measurements in the far-SOL of several tokamak experiments reveal skewed and flattened fluctuation probability density functions (PDFs), exponential auto-correlation functions and Lorenzian power spectra for positive definite variables such as ion saturation current, electron density and temperature, and gas puff imaging (GPI) intensity signals \cite{garcia-nf-2007,garcia-pop-2013,militello-ppcf-2013,kube-2016,garcia-nme-2016,theodorsen-ppcf-2016,theodorsen-nf-2017,kube-cmod,walkden-nf-2017,walkden-ppcf-2017}. Conditional averaging of large-amplitude fluctuations show that large structures exhibit fast exponential growth and slower exponential decay, with exponentially distributed peak amplitudes and exponentially distributed waiting times between them \cite{garcia-nf-2007,garcia-pop-2013,kube-2016,garcia-nme-2016,theodorsen-ppcf-2016,walkden-nf-2017,kube-cmod}. Measurements of the radial velocity is shown to have PDFs with exponential tails which are nearly symmetric around the mean value \cite{theodorsen-ppcf-2016,kube-cmod}. Previously, PDFs with exponential tails have been investigated using the so-called instanton method \cite{kim-2008,anderson-2010}. 

In order to systematize and unify these observations, a well-known reference model for intermittent fluctuations \cite{rice-1944,rice-1945,fesce-1986,kristensen-1991,jang-2004,daly-2006,narasimha-2007,elter-2015} has been introduced for SOL plasma fluctuations \cite{garcia-prl-2012,garcia-pop-2016,militello-nf-2016}. This model, called a shot noise process or filtered Poisson process (FPP), consists of a super-position of independent and identical pulse shapes with randomly distributed amplitudes, arriving according to a Poisson process. The predictions of this FPP have been shown to be in excellent agreement with experimental measurements of PDFs, auto-correlation functions and frequency power spectra, conditional averaging, and higher order statistics such as threshold level crossings and excess time statistics \cite{garcia-pop-2013,kube-2016,theodorsen-ppcf-2016,garcia-nme-2016,theodorsen-nf-2017,walkden-nf-2017,kube-cmod}. 

This model provides a framework for comparing measurements of SOL fluctuations. For example, it has been demonstrated from GPI data in Alcator C-Mod that far-SOL fluctuations have highly skewed and flattened PDFs, while PDFs close to the separatrix more closely resemble normal distributions. \cite{garcia-pop-2013,theodorsen-nf-2017} At the same time, Lorenzian power spectra with the same time scale are observed at all radial positions \cite{theodorsen-nf-2017}. Interpreting the PDFs by model, blobs are numerous and close together in the near SOL, while they are further apart in the far SOL. The blobs retain their basic shape while traveling through the SOL, however, as indicated by the universality of the spectra \cite{theodorsen-nf-2017}. By comparing PDFs and power spectra, and by estimating model parameters, measurements in the SOL of different fusion experiments, in various confinement modes and for a range of plasma parameters can be compared. We note, however, that due to the time invariance of the Poisson process, the model inherently only describes statistically stationary turbulence in the SOL.

Previous theoretical work has revealed the convergence of the lowest order moments for the process \cite{kube-2015}, extended the model to include additive noise \cite{theodorsen-ps-2017}, revealed the auto-correlation and power spectrum for different pulse shapes \cite{garcia-pop-2017-1} and for randomly distributed pulse durations \cite{garcia-pop-2017-2} and discussed the rate of threshold crossings and average time above a given threshold \cite{theodorsen-pop-2016}. It has also been demonstrated that radial propagation of filament structures with these statistical properties results in exponential profiles in the SOL, consistent with experimental observations \cite{garcia-pop-2016,militello-nf-2016}.  In this contribution, we focus on the PDF, the characteristic function and the lowest order moments of the process for various relevant amplitude distributions.

While conditional averaging has demonstrated exponentially distributed amplitudes for large-amplitude fluctuations, the statistical properties of small amplitudes has not been revealed. Assuming a positive definite time series (as is the case for ion saturation current, electron density, electron temperature or GPI intensity), exponentially distributed amplitudes is an obvious assumption. Another candidate is the Gamma distribution with shape parameter $\beta>1$. This distribution is unimodal and decays exponentially for large amplitudes, but has vanishing probability for amplitudes approaching zero. In this contribution we will compare the distributions for the FPP given exponentially and Gamma distributed amplitudes with shape parameter $\beta = 2$, since this captures the essential differences between the two distributions while allowing for analytical treatment of the PDF and the characteristic function of the FPP.

While the ion saturation current, electron density and temperature, and GPI intensity all are positive definite variables, electric potential and radial velocity are not. Thus, in order to correctly model fluctuations in these quantities, amplitude distributions with non-zero probability for negative amplitudes are required. The asymmetric Laplace distribution fulfills this requirement while still having exponential tails. The PDF of the FPP with symmetric Laplace distributed amplitudes has been favorably compared to measurements of radial velocity of filaments in the SOL \cite{theodorsen-ppcf-2016,kube-cmod}. In the asymmetric case, however, the resulting FPP does not have a closed form solution for the PDF. Thus methods for estimating model parameters based on or requiring the PDF of the process are not applicable. However, the characteristic function for the model can still be found in closed form. This allows for a method based on the empirical characteristic function, which is general enough to allow for any asymmetry in the Laplace distributed amplitudes (of which the exponential distribution is a special case) and noise level. We present this method and its application to the FPP.

\section{Moments of the filtered Poisson process}
In this section, the FPP is introduced as a model for SOL fluctuations. The general form of the moments are presented for two-sided, exponential pulse shapes and three different pulse amplitude distributions: the exponential distribution, the Gamma distribution with shape parameter $\beta = 2$ and the asymmetric Laplace distribution. We also investigate how normally distributed additive noise affects the moments of the FPP.

The FPP is given by a super-position of $K$ identical pulse shapes $\Snw$ with randomly distributed amplitudes $A_k$ arriving at times $s_k$ restricted to the range $0\leq s_k \leq T$. All random variables are assumed independent. The pulses arrive according to a Poisson process with intensity $T/\tw$, where $\tw$ is the average waiting time between pulses. Thus the arrival times are uniformly distributed on the interval $[0,T]$ and the waiting time between pulses is exponentially distributed with mean value $\tw$. 

  We express the FPP as \cite{rice-1944,rice-1945,parzen-sp,pecseli-fps}
\begin{equation}
    \Sn_K(t) = \sum_{k=1}^{K(T)} A_k \Snw\left(\frac{t-s_k}{\td}\right),
\end{equation}
where $\td$ is the fixed pulse duration time. In general, the duration times may be randomly distributed, but only the mean value of this variable (that is, $\td$) plays a role for the moments and distribution of $\Sn$, see \Appref{app:char-fun-deriv}. Thus, for simplicity of notation and without loss of generality, we will in the following consider a constant duration time. The pulse shape $\Snw(\theta)$, where $\theta$ is a unitless variable, is normalized according to
\begin{equation}
    \int_{-\infty}^\infty \rmd \theta \, \abs{\Snw(\theta)} = 1.
\end{equation}
Here and in the following, only the two-sided exponential pulse shape will be considered. This pulse shape has an asymmetry parameter $0<\lambda<1$, and is given by
\begin{equation}\label{eq:exp-wf}
    \Snw(\theta) = \begin{cases}
        \exp(\theta/\lambda), & \theta<0, \\
        \exp(-\theta/(1-\lambda)), & \theta \geq 0.
    \end{cases}
\end{equation}
We define the integral of the $n$'th power of the pulse function, which for the two-sided exponential pulse is independent of the asymmetry parameter $\lambda$,
\begin{equation}\label{eq:wf-In}
    I_n = \int_{-\infty}^\infty \rmd \theta \, \left[\Snw(\theta)\right]^n = \frac{1}{n}.
\end{equation}
While in principle $\lambda$ could be randomly distributed, as discussed in \Appref{app:char-fun-deriv}, we will in this contribution assume all pulses to be identical, with the same, fixed $\lambda$ and $\td$.

Under the assumptions given above, the characteristic function of the FPP has been derived in \Appref{app:char-fun-deriv}, and is given by \Eqref{app:char-fun-sn-end}. Inserting the integral of the $n$'th power of the pulse function given in \Eqref{eq:wf-In} into \Eqref{app:char-fun-sn-end}, the logarithm of the characteristic function of the FPP is given by the sum
\begin{equation}
  \ln C_\Sn(u) = \gamma \sum_{n=1}^\infty \frac{(i u)^n}{n!\,n} \mean{A^n},
    \label{eq:sn-char-sum}
\end{equation}
where $\gamma = \td/\tw$ is the so-called \emph{intermittency parameter} of the process. This parameter determines the degree of pulse overlap, and thereby the intermittency of the process. For low $\gamma$, each pulse duration is short compared to the average time between pulses, and the process is strongly intermittent. For high $\gamma$, many pulses arrive in the duration of one pulse event and pulse overlap becomes significant. 

The cumulants of the process are given by the coefficients in the expansion of the logarithm of the characteristic function,
\begin{equation}
  \ln C_\Sn(u) = \sum_{n=1}^\infty \kappa_n \frac{(i u)^n}{n!},
  \label{eq:cumulants_def}
\end{equation}
which according to \Eqref{eq:sn-char-sum} are given by
\begin{equation}
  \kappa_n = \frac{\gamma}{n} \mean{A^n}.
  \label{eq:sn-cumulants}
\end{equation}
The mean value of the process is $\mSn = \kappa_1$, the variance is $\rSn^2 = \kappa_2$, where rms denotes the root mean square value, and the skewness and flatness moments are related to the cumulants by \cite{rice-1944,garcia-prl-2012}
\begin{subequations} \label{eq:cumulant-to-moment}
\begin{align}
  S_\Sn &= \frac{\kappa_3}{\kappa_2^{3/2}},\label{eq:def-skew}\\
  F_\Sn &= 3+\frac{\kappa_4}{\kappa_2^2}.\label{eq:def-flat}
\end{align}
\end{subequations}
According to \Eqref{eq:sn-cumulants}, each cumulant is proportional to $\gamma$ for any amplitude distribution. Thus, the mean value is proportional to $\gamma$, the rms-value is proportional to $\gamma^{1/2}$, the skewness is proportional to $\gamma^{-1/2}$ and the flatness to $1/\gamma$. For increasing $\gamma$, any FPP will have mean and rms tending to infinity (for finite $I_n$ and $\left< A^n \right>$) and vanishing skewness and flatness. For the FPP there is a universal parabolic relationship between $F_\Sn$ and $S_\Sn$, independent of the intermittency parameter $\gamma$ \cite{garcia-prl-2012,garcia-pop-2016}
\begin{equation}
  F_\Sn = 3 + \frac{\kappa_2 \kappa_4}{\kappa_3^2} S_\Sn^2 = 3 + \frac{I_2 I_4}{I_3^2} \frac{\mean{A^2} \mean{A^4}}{\mean{A^3}^2}S_\Sn^2 = 3 + \frac{9}{8} \frac{\mean{A^2} \mean{A^4}}{\mean{A^3}^2}S_\Sn^2.
  \label{eq:parabolic_relation_general}
\end{equation}
The physical basis for a parabolic relationship between skewness and kurtosis has been explored previously \cite{sattin-2009,bergsaker-2015}. We note that while many relationships between skewness and kurtosis based on \Eqref{eq:cumulant-to-moment} are possible, only the one presented in \Eqref{eq:parabolic_relation_general} is independent of the intermittency parameter $\gamma$.

We will consider 3 different amplitude distributions for the FPP; the exponential distribution, the Gamma distribution with shape parameter $\beta = 2$ and the asymmetric Laplace distribution. These all give closed form expressions for the characteristic function of $\Sn$. The PDF of the exponential distribution has a finite value for $A=0$ and is monotonically decreasing. The Gamma distribution with shape parameter $2$ is unimodal and tends to $0$ for $A \to 0$. Since both have exponential tails, comparing the distribution of the FPP with exponentially and Gamma distributed amplitudes will highlight the importance of small-amplitude pulses while keeping the effect of large-amplitude pulses equal. The Laplace distribution allows for both positive and negative values of $A$, whereas the exponentially and Gamma distributed amplitudes are strictly positive. Thus the Laplace distribution is the only one of these capable of describing measurement data which is not positive definite. 

\subsection{Exponentially distributed amplitudes}
The exponential distribution is a one parameter distribution with scale parameter $\alpha>0$. The distribution and its moments are given by
\begin{subequations}\label{eq:exp-dist}
\begin{align}
    P_A(A;\alpha) &= \frac{1}{\alpha} \exp(-A/\alpha),\,A>0,\label{eq:exp-dist-dist}\\
    \mean{A^n} &= \alpha^{n} n!,\label{eq:exp-dist-mom}
\end{align}
\end{subequations}
for integer values of $n$.

For exponentially distributed amplitudes, the first four moments of $\Sn$ are given by \cite{garcia-prl-2012,garcia-pop-2016}
\begin{subequations}
\begin{align}
    \mSn &= \gamma \alpha, \\
    \rSn^2 &= \gamma \alpha^2, \\
    S_{\Sn} &= \frac{2}{\gamma^{1/2}}, \\
    F_{\Sn} &= \frac{6}{\gamma} + 3,
\end{align}
\end{subequations}
and we have the parabolic relationship between the skewness and flatness moments,
\begin{equation}
  F_\Sn = 3+\frac{3}{2}S_\Sn^2,
  \label{eq:parabolic-sn-exp}
\end{equation}
where the pre-factor is simply $3/2$, as the scale parameter is cancelled out.

\subsection{Gamma distributed amplitudes}
The Gamma distribution has a shape parameter $\beta>0$ and a scale parameter $\alpha>0$. It is given by
\begin{subequations}\label{eq:gamma-dist}
\begin{align}
  P_A(A;\alpha,\beta) &= \frac{A^{\beta-1}}{\alpha^{\beta} \Gamma(\beta)}\exp(-A/\alpha),\, A>0, \label{eq:gamma-dist-dist}\\
  \mean{A^n} &= \alpha^n \frac{\Gamma(\beta+n)}{\Gamma(\beta)}.\label{eq:gamma-dist-mom}
\end{align}
\end{subequations}
For $\beta = 1$, this is equivalent to the exponential distribution with scale parameter $\alpha$. For $\beta = 2$, we have 
\begin{subequations}\label{eq:gamma-dist-b2}
\begin{align}
  P_A(A;\alpha,\beta) &= \frac{A}{\alpha^{2}}\exp(-A/\alpha),\, A>0, \label{eq:gamma-dist-b2-dist}\\
  \mean{A^n} &= \alpha^n \Gamma(2+n).\label{eq:gamma-dist-b2-mom}
\end{align}
\end{subequations}
For large amplitudes $A$, the Gamma distribution has an exponential tail with the same decay rate as the exponential distribution with equal $\alpha$. The moments are not equal, however, as $\Gamma(2+n) = (1+n)!$, giving a factor $(1+n)$ more for all moments in the case of Gamma distributed pulse amplitudes.

For Gamma distributed amplitudes with general shape parameter $\beta$, the first four moments are given by
\begin{subequations}\label{eq:sn-moments-gamma-amp}
\begin{align}
    \mSn &= \gamma \alpha \beta, \\
    \rSn^2 &= \gamma \alpha^2 \frac{\beta (\beta+1)}{2}, \\
    S_{\Sn} &= \frac{2^{3/2}}{ 3 \gamma^{1/2}} \frac{\beta+2}{\left[ \beta \left( \beta+1 \right) \right]^{1/2}}, \\
    F_{\Sn} &= \frac{1}{\gamma} \frac{(\beta+2)(\beta+3)}{\beta (\beta+1)} + 3.
\end{align}
\end{subequations}
The parabolic relationship between the skewness and flatness moments depends on the shape parameter for the amplitude distribution,
\begin{equation}
  F_\Sn = 3+\frac{9}{8} \frac{\beta+3}{\beta+2} S_\Sn^2.
  \label{eq:parabolic-sn-gamma}
\end{equation}
Setting $\beta = 1$ gives the same moments and parabolic relation as for the exponentially distributed amplitudes. For the special case $\beta=2$, the moments are
\begin{subequations}\label{eq:sn-moments-gamma-amp-b2}
\begin{align}
    \mSn &= 2 \gamma \alpha, \\
    \rSn^2 &= 3 \gamma \alpha^2, \\
    S_{\Sn} &= \frac{8}{ 3^{3/2} \gamma^{1/2}}, \\
    F_{\Sn} &= \frac{10}{3 \gamma} + 3.
\end{align}
\end{subequations}
and the parabolic relationship simplifies to
\begin{equation}
  F_\Sn = 3+\frac{45}{32}S_\Sn^2.
  \label{eq:parabolic-sn-gamma-b2}
\end{equation}
This relationship is very close to the case of exponentially distributed amplitudes, \Eqref{eq:parabolic-sn-exp}, (it would be equal for the prefactor $48/32 = 3/2$).

\subsection{Asymmetrically Laplace distributed amplitudes}
The asymmetric Laplace distribution can be formulated in a few different ways, see for example \Refs{kozubowski-2000,reed-2006}. We will use a different formulation which easily admits the exponential distribution as a limiting case. With $\alpha>0$ as a scale parameter and $0<\beta<1$ as a shape parameter, we have
\begin{subequations}\label{eq:laplace-dist}
\begin{align}
P_A(A;\alpha,\beta) &= \frac{1}{ 2 \alpha} \begin{cases} \exp\left( - \frac{A}{2 \alpha (1-\beta)} \right), & A>0, \\ \exp\left( \frac{A}{2 \alpha \beta} \right), & A<0, \end{cases}\label{eq:laplace-dist-dist}\\
\mean{A^n} &= (2 \alpha)^n n! \left[ (-1)^n \beta^{n+1} + (1-\beta)^{n+1}\right].\label{eq:laplace-dist-mom}
\end{align}
\end{subequations}
This distribution is symmetric for $\beta = 1/2$, and is equivalent to the exponential distribution in the limit $\beta \to 0$. In the limit $\beta \to 1$, the distribution is an exponential distribution mirrored around $A=0$, with zero probability for positive $A$-values and finite probability for negative $A$-values. 

For Laplace distributed pulse amplitudes, the first four moments are given by
\begin{subequations}
\begin{align}
    \mSn &= 2 \gamma \alpha (1-2\beta), \\
    \rSn^2 &= 4 \gamma \alpha^2 [\beta^3+(1-\beta)^3], \\
    S_{\Sn} &=  \frac{(1-\beta)^4-\beta^4}{ [\beta^3+(1-\beta)^3]^{3/2}}\frac{2}{\gamma^{1/2}}, \\
    F_{\Sn} &= \frac{[\beta^5+(1-\beta)^5]}{[\beta^3+(1-\beta)^3]^2}\frac{6}{\gamma} + 3.
\end{align}
\end{subequations}
and the parabolic relationship between skewness and flatness is,
\begin{equation}
  F_\Sn = 3+\frac{3}{2}\frac{[\beta^5+(1-\beta)^5][\beta^3+(1-\beta)^3]}{[(1-\beta)^4-\beta^4]^2} S_\Sn^2.
  \label{eq:parabolic-sn-Laplace}
\end{equation}
For $\beta = 0$, we have the same expressions as for the exponentially distributed amplitudes. For $\beta = 1/2$, the Laplace distribution is symmetric, so all odd moments of $A$ vanish, see \Eqref{eq:laplace-dist-mom}, giving
\begin{subequations}
\begin{align}
    \mSn &= 0, \\
    \rSn^2 &= \gamma \alpha^2, \\
    S_{\Sn} &=  0, \\
    F_{\Sn} &= \frac{6}{\gamma} + 3.
\end{align}
\end{subequations}
In this case, there is no parabolic relationship between skewness and flatness, since $S_{\Sn} = 0$. As $\beta$ approaches $1/2$ (from either side), the prefactor in \Eqref{eq:parabolic-sn-Laplace} tends to infinity.

\subsection{Comparisons}

\begin{figure}
  \centering
  \subfloat{\includegraphics[width = 0.47\textwidth]{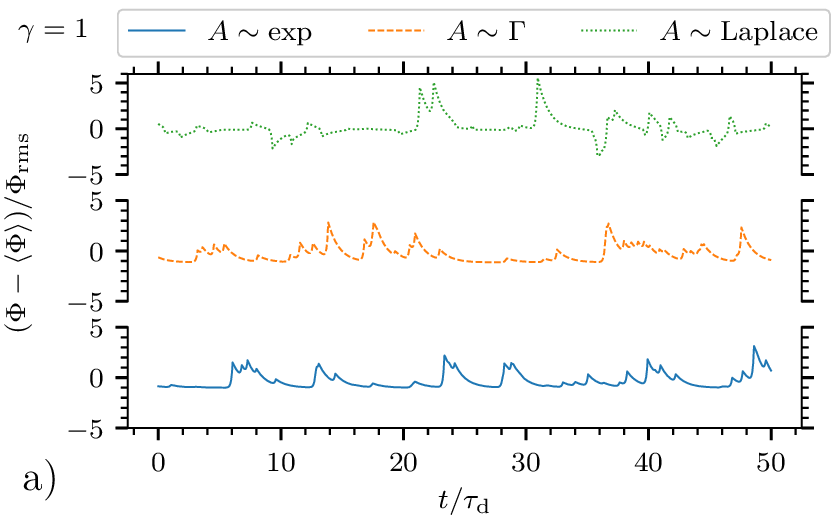}\label{fig:sn-a-var-g1}}
  ~
  \subfloat{\includegraphics[width = 0.47\textwidth]{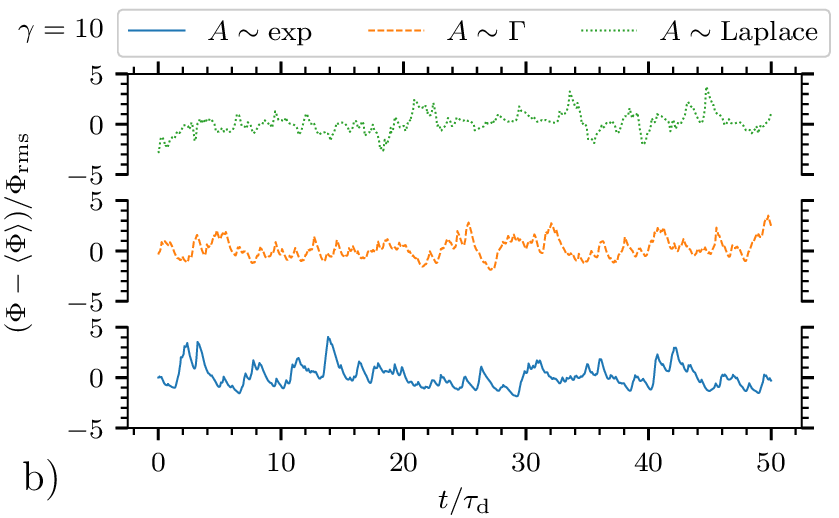}\label{fig:sn-a-var-g10}}
  \caption{\label{fig:sn-a-var} Realizations of the filtered Poisson process for various pulse amplitude distributions and values of the intermittency parameter $\gamma$. The pulse asymmetry parameter $\lambda$ is $1/10$.}
\end{figure}

In \Figref{fig:sn-a-var}, realizations of the process are presented for $\gamma \in \{1,10\}$ and the amplitude distributions given above. The bottom (blue) lines give realizations with amplitudes distributed according to an exponential distribution. The middle (orange dashed) lines are for Gamma distributed amplitudes with shape parameter $\beta = 2$, and the top (green dotted) realizations are computed using the Laplace distribution with shape parameter $\beta = 1/2$. In all cases, the realizations have been normalized to have zero mean and unit standard deviation, in order to remove the dependency on the scale parameter in the amplitude distribution. In \Figref{fig:sn-a-var-g1}, $\gamma = 1$ and all processes are strongly intermittent, alternating between periods of activity and inactivity. In \Figref{fig:sn-a-var-g10}, $\gamma = 10$, and the large degree of pulse overlap leads to weaker intermittency and makes individual pulses harder to discern. While the signals with exponentially and Gamma distributed amplitudes are easy to separate visually from the Laplace case for $\gamma = 1$, this is not true for the case with $\gamma = 10$, where all realizations look like a random walk around a mean value. Indeed, it can be shown that the PDF of the normalized process $\nSn = (\Sn-\mSn)/\rSn$ approaches a standard normal distribution as $\gamma \to \infty$, independent of the amplitude distribution and the pulse shape \cite{rice-1944,pecseli-fps}. Visually, it is very difficult to separate the filtered Poisson process with exponentially distributed amplitudes from the one with Gamma distributed amplitudes with shape parameter $\beta=2$.

\begin{figure}
  \centering
    \includegraphics[width = \textwidth]{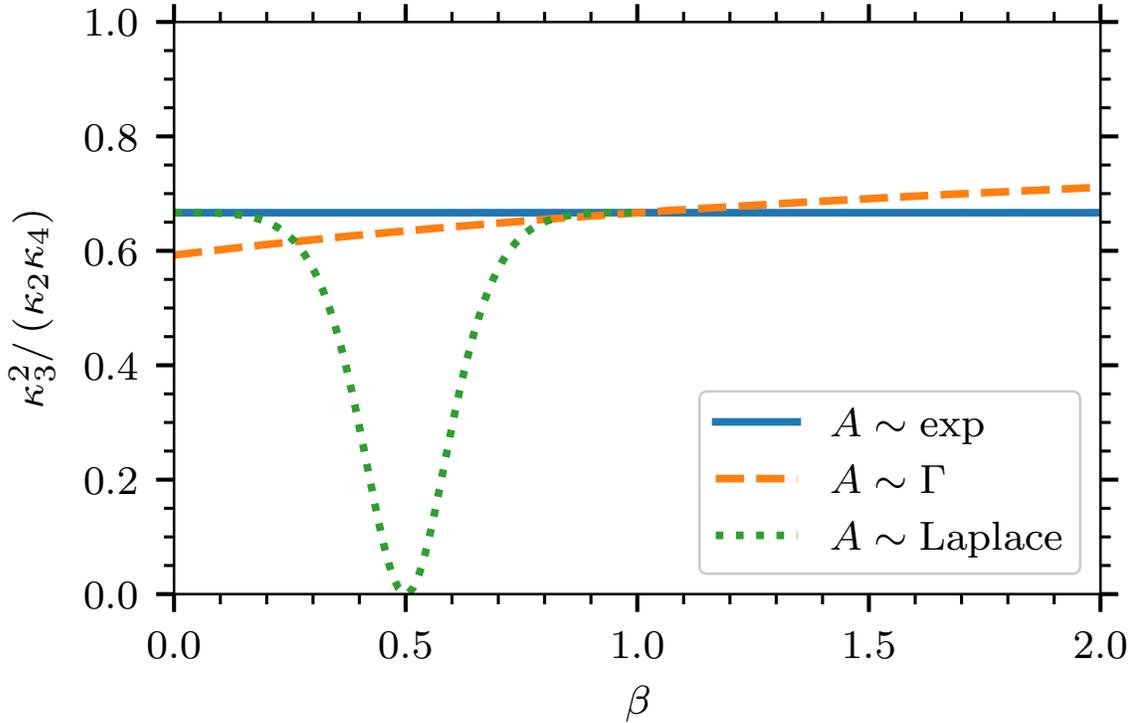}
  \caption{\label{fig:parabolic-relationship-prefac} The inverse of the prefactor in the parabolic relationship between the skewness and flatness moments for the filtered Poisson process with exponentially (blue), Gamma (orange dashed) and Laplace (green dotted) distributed pulse amplitudes as function of the shape parameter of the amplitude distribution.}
\end{figure}

In \Figref{fig:parabolic-relationship-prefac}, the inverse of the prefactor in the parabolic relationship between the skewness and flatness moments as a function of $\beta$ is shown for exponentially distributed amplitudes (blue), Gamma distributed amplitudes (orange dashed) and Laplace distributed amplitudes (green dotted). The inverse is used, since the prefactor itself tends to infinity in the case of Laplace distributed amplitudes for $\beta \to 1/2$, as discussed above. This prefactor is constant for exponentially distributed amplitudes, since the exponential distribution has no shape parameter. For Gamma distributed amplitudes, the inverse of the prefactor is smaller than for exponentially distributed amplitudes for $\beta<1$ and larger for $\beta>1$, thus the prefactor itself is larger for $\beta<1$ and smaller for $\beta>1$. From \Eqref{eq:parabolic-sn-gamma} we see that the prefactor approaches $9/8$ as $\beta \to \infty$. The FPP with Laplace distributed amplitudes and $\beta = 0$ or $\beta = 1$ has the same prefactor in the parabolic relation as the FPP with exponentially distributed amplitudes. The skewness is equal in magnitude but with different sign for these two cases, giving the same prefactor in the  parabolic relationship.

\subsection{Additive noise}
In many applications of the model, there may be some normally distributed additive noise to the process, either as white noise connected to measurements or as noise with the same power spectrum as the FPP, connected to the dynamics. This situation has been explored in detail for exponentially distributed amplitudes and one-sided exponential pulses \cite{theodorsen-ps-2017}. We assume the noise process, denoted by $X$, to be normally distributed and independent of the FPP, and denote the signal with additive noise as 
\begin{equation}
  \Snn = \Sn + X.
  \label{eq:sn-with-noise-def}
\end{equation}
The basic properties of the distribution of a sum of independent random variables are reviewed in \Appref{app:sum-rv}. For a normal distribution, only the first two cumulants are non-zero. Using that the variance of the process is given by the second cumulant, we define the noise ratio as
\begin{equation}
  \epsilon = \frac{X_\rms^2}{\rSn^2} = \frac{\kappa_2^X}{\kappa_2^\Sn}.
  \label{eq:sn-noise-ratio}
\end{equation}
Using this noise ratio and that the cumulants of a sum of independent random variables is the sum of their cumulants, see \Eqref{app:sum-rv-cumulants}, we have
\begin{subequations}\label{eq:sn-noise-moments}
\begin{align}
  \mSnn &= \kappa^\Sn_1+\kappa^X_1,\\
  \rSnn^2 &= (1+\epsilon) \kappa^\Sn_2,\\
  S_\Snn &= \frac{\kappa^\Sn_3}{\left[ (1+\epsilon) \kappa^\Sn_2\right]^{3/2}},\\
F_\Snn &= 3 + \frac{\kappa^\Sn_4}{\left[ (1+\epsilon) \kappa^\Sn_2 \right]^2}.
\end{align}
\end{subequations}
The parabolic relationship between the skewness and flatness moments is 
\begin{equation}
  F_\Snn = 3 + \frac{\kappa_4^\Sn \left[ (1+\epsilon) \kappa_2^\Sn  \right]}{\left( \kappa_3^\Sn \right)^2} S_\Snn^2.
  \label{eq:sn-noise-parabolic}
\end{equation}
The effect of noise on the moments is to increase the variance and decrease the skewness and flatness, leading to $\Snn$ more closely resembling a normally distributed process than $\Sn$. As $\epsilon$ increases, the prefactor in the parabolic relationship increases as well.

\section{Probability distributions}
Under the assumptions that we have an FPP with fixed $\td$ and $\lambda$ and independent amplitudes and arrival times, the characteristic function of the FPP has been derived in \Appref{app:char-fun-deriv}, and is given in various forms by \Eqsref{app:char-fun-sn-integral-start}, \eqref{app:char-fun-sn-integral} and \eqref{app:char-fun-sn-end}. For the two-sided exponential pulse function, we can split the integral in \Eqref{app:char-fun-sn-integral} into two parts, one over negative values of $\theta$ and one over positive values of $\theta$, and substitute the integration for the pulse shape in \Eqref{eq:exp-wf}. This gives
\begin{equation}
  \ln C_\Sn(u) = \gamma \int\limits_0^u \rmd v \, \frac{C_A(v)-1}{v},
  \label{eq:sn-char-int}
\end{equation}
where $C_A$ is the characteristic function of the amplitudes. Inserting the integral of the $n$'th power of the pulse function given in \Eqref{eq:wf-In} into \Eqref{app:char-fun-sn-end}, we can alternatively give the characteristic function of the FPP as the sum
\begin{equation}
  \ln C_\Sn(u) = \gamma \sum_{n=1}^\infty \frac{(i u)^n}{n!\,n} \mean{A^n},
    \label{eq:sn-char-sum}
\end{equation}
as in the previous section. The PDF of $\Sn$ is given by
\begin{equation}
    P_\Sn(\Sn) = \frac{1}{2 \pi} \int_{-\infty}^\infty \rmd u \, \exp\left( -i\Sn u\right) C_\Sn(u).
    \label{eq:p-sn-from-char}
\end{equation}
We will in the following frequently use the normalization
\begin{equation}
  \nSn = \frac{\Sn-\mSn}{\rSn},
  \label{eq:norm-def}
\end{equation}
which removes the dependence on the scale parameter in the amplitude distribution. The PDF of $\nSn$ is in general
\begin{equation}
  P_\nSn(\nSn) = \rSn P_\Sn(\rSn \nSn + \mSn),
  \label{eq:norm-sn-pdf}
\end{equation}
while its characteristic function is
\begin{equation}
  C_\nSn(u) = \exp\left( -i \frac{\mSn}{\rSn} u \right) C_\Sn\left( \frac{u}{\rSn} \right).
  \label{eq:norm-sn-cf}
\end{equation}

\subsection{Exponential amplitude distribution}
\begin{figure}
  \centering
    \includegraphics[width = \textwidth]{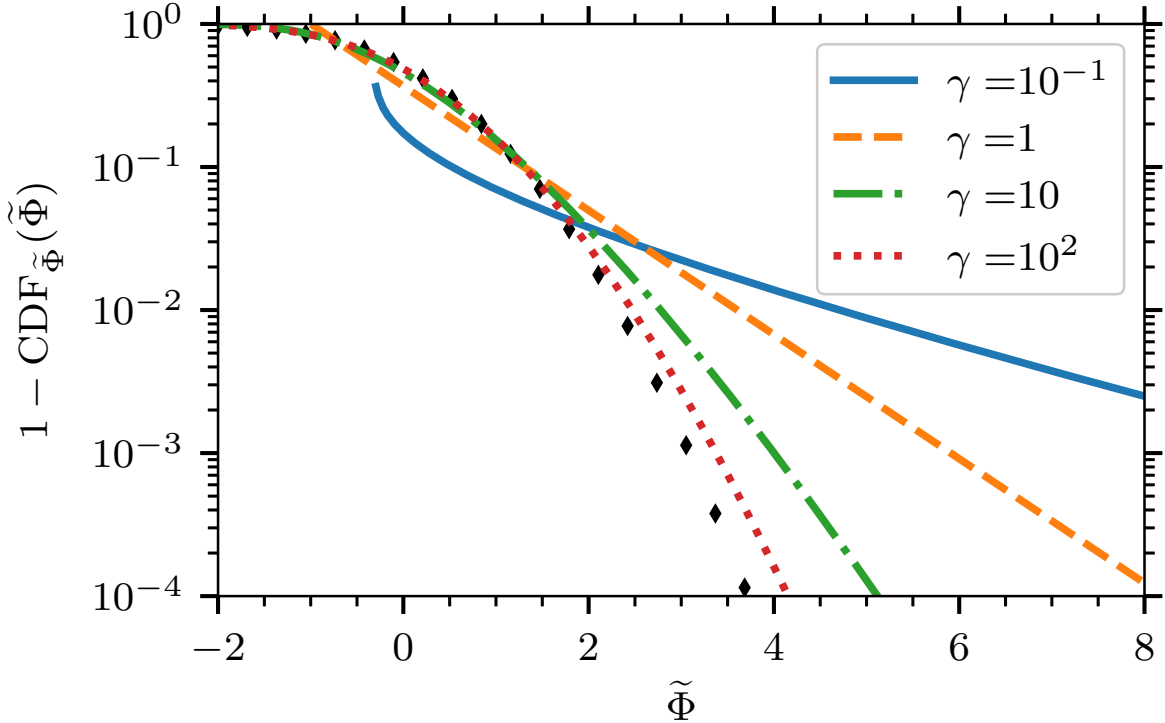}
  \caption{\label{fig:ccdf-exp-amp} Complementary cumulative distribution function of the normalized filtered Poisson process for various values of the intermittency parameter $\gamma$. The black diamonds indicate a normal distribution with vanishing mean and unit standard deviation.}
\end{figure}

In the case of exponentially distributed amplitudes, it is well known \cite{bondesson-1982,pecseli-fps,daly-2006,daly-2010,garcia-prl-2012} that the distribution of $\Sn$ is a gamma distribution with shape parameter $\gamma$ and scale parameter $\alpha$,
\begin{subequations}\label{eq:sn-exp-amp-dist}
\begin{align}
  P_\Sn(\Sn) &= \frac{\Sn^{\gamma-1}}{\alpha^\gamma \Gamma(\gamma)}\exp\left(-\frac{\Sn}{\alpha} \right),\,\Sn>0, \label{eq:sn-exp-amp-pdf} \\
  C_\Sn(u) &= \left( 1-i \alpha u \right)^{-\gamma}.
  \label{eq:sn-exp-amp-cf}
\end{align}
\end{subequations}
Using the normalization in \Eqref{eq:norm-def}, the dependence explicit on $\alpha$ disappears, and the distribution becomes
\begin{subequations}\label{eq:nsn-exp-amp-dist}
\begin{align}
  P_{\nSn}(\nSn) &= \frac{\gamma^{\gamma/2}}{\Gamma(\gamma)} (\nSn + \gamma^{1/2})^{\gamma-1} \exp( -\gamma^{1/2} \nSn - \gamma  ), \label{eq:nsn-exp-amp-pdf}\\
  C_{\nSn}(u) &= \exp\left( -i \sqrt{\gamma}u \right)\left( 1-i \frac{u}{\sqrt{\gamma}} \right)^{-\gamma}.
  \label{eq:nsn-exp-amp-cf}
\end{align}
\end{subequations}
In \Figref{fig:ccdf-exp-amp}, the effect of the parameter $\gamma$ is illustrated by presenting the complementary cumulative distribution function (cCDF) of $\nSn$ for various parameter values. The cCDF at a given function value $\nsn$ can be interpreted as the probability that the random variable takes the value $\nsn$ or a larger value. It can also be interpreted as the fraction of time a signal spends above a threshold value $\nsn$. $\Sn$ is positive definite, so the lowest possible value for $\nSn$ is $\nSn=-\sqrt{\gamma}$. For small $\gamma$, the cCDF falls off slowly with increasing $\nSn$, indicating high probability of large amplitude fluctuations. As $\gamma$ increases, the probability of large values of $\nSn$ decreases, as the signal transitions from an intermittent signal to one resembling random motion around a mean value. As stated earlier, for $\gamma \to \infty$, the process $\nSn$ approaches a standard normal distribution, presented by the black diamonds in \Figref{fig:ccdf-exp-amp}. For the $\gamma$-values presented here, the Gamma-distributed signals all have higher probability of fluctuations with amplitude $\nSn > 2$ compared to a normally distributed signal, highlighting the importance of intermittency for threshold phenomena.

\subsection{Gamma amplitude distribution}\label{sec:sn-gamma-amp}
\begin{figure}
  \centering
    \includegraphics[width = \textwidth]{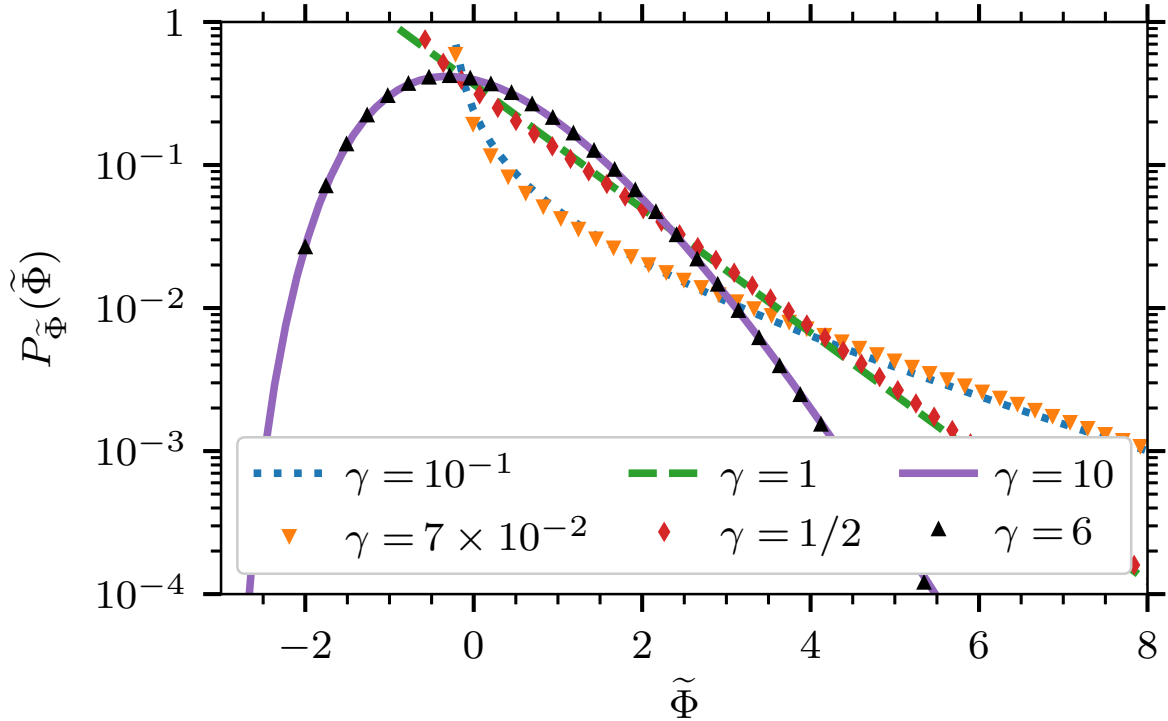}
  \caption{\label{fig:pdf-compare-exp-gamma} Probability density functions for the filtered Poisson process with exponential amplitude distribution (full lines) and Gamma amplitude distribution with shape parameter $\beta=2$ (dashed lines). Both are given for various values of the intermittency parameter $\gamma$.}
\end{figure}

Using gamma distributed amplitudes, we have 
\begin{equation}
  \ln C_\Sn(u) = \gamma \beta \, i \alpha u \, {_3}F_2\left[ \begin{matrix} 1&1&1+\beta\\2&2& \end{matrix}; i \alpha u\right],
  \label{eq:cf-gen-sn-gamma-amp}
\end{equation}
where ${_p}F_q$ is the generalized hypergeometric function \cite{nist-dlmf}. For $\beta=1$, this simplifies to the characteristic function for an exponential amplitude distribution, as discussed above. For $\beta=2$, the characteristic function is
\begin{equation}
  \ln C_\Sn(u) = \gamma \left[ \frac{1}{1-i \alpha u} -1 -\ln(1-i \alpha u) \right],
  \label{eq:log-cf-k2-sn-gamma-amp}
\end{equation}
giving
\begin{equation}
  C_\Sn(u) = \exp(-\gamma) \exp\left(\frac{\gamma}{1-i \alpha u} \right) (1-i \alpha u)^{-\gamma}.
  \label{eq:cf-k2-sn-gamma-amp}
\end{equation}
In \Appref{app:sn-gamma-amp-cf-to-pdf}, the corresponding PDF is shown to be \cite{daly-2006,daly-2010}
\begin{equation}
  P_\Sn(\Sn) = \frac{1}{\alpha} \left( \frac{\Sn}{\gamma \alpha} \right)^{(\gamma-1)/2} \exp\left(-\gamma-\frac{\Sn}{\alpha}\right)  I_{\gamma-1} \left( 2 \sqrt{\gamma \frac{\Sn}{\alpha}} \right).
  \label{eq:pdf-sn-gamma-amp}
\end{equation}
Here, $I$ is the modified Bessel function of the first kind \cite{nist-dlmf}. Using the mean and rms-values given by \Eqref{eq:sn-moments-gamma-amp-b2}, we have that the distribution of the normalized variable $\nSn = (\Sn-\mSn)/\rSn$ is again independent of $\alpha$, and we have
\begin{subequations}\label{eq:nsn-gamma-amp-dist}
\begin{align}
  P_\nSn(\nSn) &= \sqrt{3\gamma} \left( \sqrt{\frac{3}{\gamma}}\nSn+2 \right)^{(\gamma-1)/2} \exp\left(-\sqrt{3\gamma}\nSn-3\gamma \right) I_{\gamma-1} \left( 2\gamma \sqrt{\sqrt{\frac{3}{\gamma}}\nSn+2} \right), \label{eq:nsn-gamma-amp-pdf}\\
  C_{\nSn}(u) &= \exp\left( -\gamma i \frac{u}{\sqrt{3 \gamma}} \frac{1+2 i u/\sqrt{3 \gamma}}{1-i u /\sqrt{3 \gamma}} \right)\left( 1-i \frac{u}{\sqrt{ 3 \gamma}} \right)^{-\gamma}. \label{eq:nsn-gamma-amp-cf}
\end{align}
\end{subequations}
A comparison between the PDF in \Eqref{eq:nsn-gamma-amp-pdf} and the PDF for the case of exponentially distributed amplitudes, given by \Eqref{eq:nsn-exp-amp-pdf}, is presented in \Figref{fig:pdf-compare-exp-gamma} for various values of the intermittency parameter. It is evident that the PDF of the FPP with Gamma distributed amplitudes can be very well approximated by the PDF of a FPP with exponentially distributed amplitudes and slightly larger $\gamma$. Thus the PDF seems a poor choice for differentiating Gamma distributed amplitudes with $\beta = 2$ from exponentially distributed amplitudes in a given realization of the process.

\subsection{Asymmetric Laplace amplitude distribution}
\begin{figure}
  \centering
  \subfloat{
    \includegraphics[width = 0.47\textwidth]{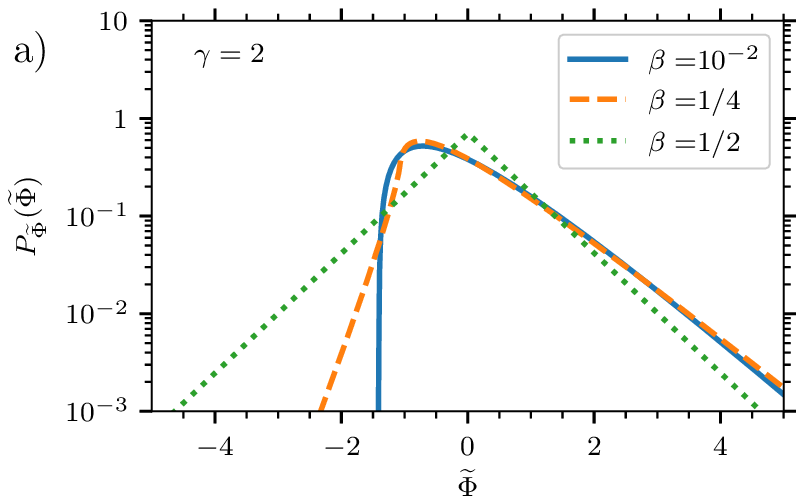}
    \label{fig:pdf-asymm-laplace-amp-bchange}}
  ~
  \subfloat{
    \includegraphics[width = 0.47\textwidth]{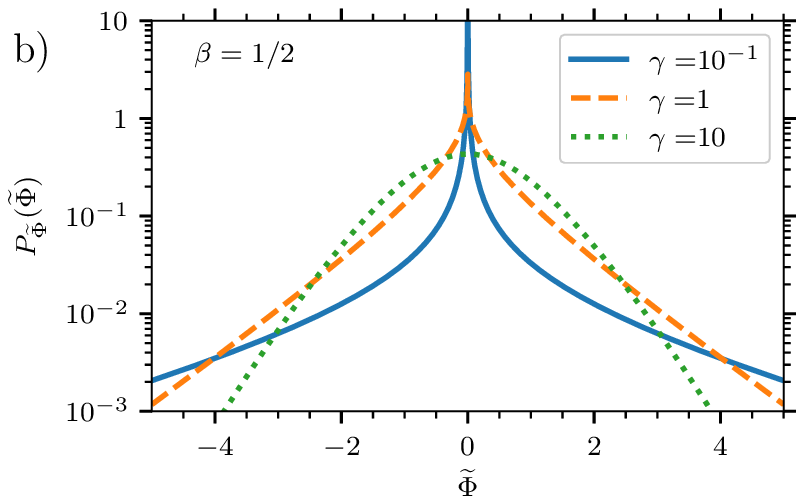}
    \label{fig:pdf-asymm-laplace-amp-gchange}}
  \caption{\label{fig:pdf-asymm-laplace-amp} Probability distribution functions of the normalized filtered Poisson process with Laplace distributed amplitudes for various $\gamma$ and $\beta$.}
\end{figure}

With asymmetrically Laplace distributed amplitudes, the characteristic function of the filtered Poisson process is
\begin{equation}\label{eq:sn-al-char}
    C_\Sn(u) = (1+i\, 2 \alpha \beta u)^{-\gamma \beta} (1-i\, 2 \alpha (1-\beta) u)^{-\gamma (1-\beta)}.
\end{equation}
Note that this can be seen as the characteristic function of a sum of two independent gamma distributed variables, one over positive values with shape parameter $\gamma (1-\beta)$ and scale parameter $2 \alpha (1-\beta)$, and one over negative values with shape parameter $\gamma \beta$ and scale parameter $2 \alpha \beta$.  While the PDF is in general not possible to find in closed form, a numerical estimate can be found by noting that the PDF of a sum of independent random variables is the convolution of their respective distributions. Estimating the two gamma distributions and convolving them numerically gives PDFs as illustrated in \Figref{fig:pdf-asymm-laplace-amp}. In \Figref{fig:pdf-asymm-laplace-amp-bchange}, the intermittency parameter is $\gamma =2$ and the shape parameter $\beta$ varies. For $\beta=1/2$, this is a symmetric Laplace distribution. As $\beta \to 0$, the PDF approaches a Gamma distribution. \Figref{fig:pdf-asymm-laplace-amp-gchange} shows the distribution for $\beta = 1/2$ and various values of $\gamma$. As discussed below, one can find the PDF in closed form in this case. The PDF is symmetric around $\nSn = 0$ for all $\gamma$. For small $\gamma$, it is sharply peaked and convex around $\nSn = 0$, while for large $\gamma$, it is concave and approaches a normal distribution as $\gamma$ increases. For all combinations of $\beta$ and (finite) $\gamma$, the PDF has exponential tails.

In the limit $\beta \to 0$, the Laplace distribution approaches the exponential distribution and the standard Gamma distribution for $\Sn$ is recovered. In the case $\beta = 1/2$, the Laplace distribution is symmetric and we can find the PDF in closed form. The distribution is \cite{daly-2006,daly-2010,theodorsen-ppcf-2016,garcia-pop-2017-2}
\begin{subequations}\label{eq:sn-symm-laplace-amp-dist}
  \begin{align}
    P_{\Sn}(\Sn;\gamma,\alpha,\beta = 1/2) &= \frac{1}{\pi^{1/2} \alpha \Gamma(\gamma/2)} \left( \frac{\abs{\Sn}}{2 \alpha} \right)^{(\gamma-1)/2} K_{(\gamma-1)/2}\left( \frac{\abs{\Sn}}{\alpha} \right), \label{eq:sn-symm-laplace-amp-pdf} \\
    C_\Sn(u) &= \left( 1 + \alpha^2 u^2 \right)^{-\gamma/2}, \label{eq:sn-symm-laplace-amp-cf}
  \end{align}
\end{subequations}
where $K$ is the modified Bessel function of the second kind \cite{nist-dlmf}. The normalized variable $\nSn$ has the distribution
\begin{subequations}\label{eq:nsn-symm-laplace-amp-dist}
  \begin{align}
    P_{\nSn}(\nSn;\gamma,\beta = 1/2) &= \frac{\gamma^{1/2}}{\pi^{1/2} \Gamma(\gamma/2)} \left( \frac{\gamma^{1/2} \abs{\nSn}}{2} \right)^{(\gamma-1)/2} K_{(\gamma-1)/2}\left(\gamma^{1/2} \abs{\nSn} \right), \label{eq:nsn-symm-laplace-amp-pdf}\\
    C_\nSn(u) &= \left( 1 + \frac{u^2}{\gamma} \right)^{-\gamma/2}.\label{eq:nsn-symm-laplace-amp-cf}
  \end{align}
\end{subequations}
This PDF is presented in \Figref{fig:pdf-asymm-laplace-amp-gchange} for various values of the intermittency parameter $\gamma$.

\subsection{Additive noise}\label{sec:cf-most-general-fpp}
Adding noise to the FPP is straightforward as long as only the characteristic function is considered. Using the FPP with asymmetrically Laplace distributed amplitudes and additive noise as an example, we have
\begin{equation}
    \Psi(t) = \Phi(t) + X(t),
\end{equation}
where $\Phi$ is a FPP with Laplace distributed amplitudes and $X$ is normally distributed noise with vanishing mean and standard deviation $X_\rms$. The characteristic function of $\Psi$ is the product of the characteristic functions of $\Sn$ and $X$, see \Appref{app:sum-rv}. We have 
\begin{equation}
    C_\Snn(u) = \left(1+ i 2 \beta \alpha u\right)^{-\gamma \beta} (1 - i 2(1-\beta)\alpha u)^{-\gamma (1-\beta)} \exp\left(-\frac{1}{2} X_\rms^2 u^2 \right).
\end{equation}
Using the noise parameter $\epsilon$ from \Eqref{eq:sn-noise-ratio}, we have $X_\rms^2 = \epsilon \Phi_\rms^2$. The moments for the FPP with additive noise from \Eqref{eq:sn-noise-moments} give $\tmean{\Psi} = 2 \alpha \gamma (1-2 \beta)$ and $\Psi_\rms^2 = (1+\epsilon)\Phi_\rms^2 = 4 \alpha^2 \gamma (1+\epsilon)\left[ \beta^3+(1-\beta)^3 \right] $. Normalizing the process by $\nSnn = (\Snn -\mSnn)/\rSnn$ and using \Eqref{eq:norm-sn-cf}, we have
\begin{multline}
    C_{\nSnn}(u) = \left(1+ i \frac{\beta u}{\sqrt{\gamma (1+\epsilon)} B(\beta)}\right)^{-\gamma \beta} \left(1 - i \frac{(1-\beta) u}{\sqrt{\gamma (1+\epsilon)} B(\beta)}\right)^{-\gamma (1-\beta)} \\ \exp\left(-\frac{\epsilon u^2 }{2(1+\epsilon)} - i \frac{\sqrt{\gamma} (1-2\beta) u}{\sqrt{1+\epsilon}B(\beta)} \right),
\end{multline}
where $B(\beta) = \sqrt{(1-\beta)^3+\beta^3}$. Performing the inverse Fourier transform on this expression to get the PDF of $\nSnn$ does in general not lead to a closed analytical expression. However, it can be done for the FPP with exponentially distributed amplitudes (that is, in the limit $\beta \to 0$) \cite{theodorsen-ps-2017}.

\section{Parameter estimation}
In \Secref{sec:sn-gamma-amp}, it was shown that the distribution of an FPP with Gamma distributed amplitudes with shape parameter $\beta =2$ does not differ significantly from an FPP with exponentially distributed amplitudes and a slightly larger $\gamma$. The exponential distribution is a special case of the asymmetric Laplace distribution, and so we consider the FPP with asymmetrically Laplace distributed amplitudes to describe the most general form of the distribution of the FPP presented in this contribution. Adding noise as well gives the process described in \Secref{sec:cf-most-general-fpp}. This process does not in general have a closed form PDF, and so standard methods for estimating the process parameters which rely on the PDF does not work. We do, however, have a closed form for the characteristic function. Estimating parameters using the characteristic function has been discussed in \Refs{feuerverger-1977,feuerverger-1981,yu-2004}. The main problem is finding a reasonable set of variables for the characteristic function. This has been considered by \Refs{tran-1998,zhang-2016}. All consider dividing the characteristic function into real and imaginary components explicitly. The approach taken here is more compact, and should be equivalent.

We have a set of independent and identically distributed observations $Y_1,Y_2,\dots,Y_N$. We assume we know the distribution family, but not the parameters; the PDF is given by $P(y;\bm{\theta})$ with the corresponding characteristic function $C(u,\bm{\theta})$, where $\bm{\theta}$ is a vector of parameters. The goal is to estimate these parameters from the observations.
Define the empirical characteristic function
\begin{equation}
    C_N(u) = \frac{1}{N} \sum_{n=1}^N \exp\left(i u Y_n\right),
\end{equation}
and the error function
\begin{equation}
    \varepsilon_N(u,\bm{\theta}) = N^{1/2} \left( C_N(u) - C(u,\bm{\theta}) \right).
\end{equation}
Given a discrete set of sampling points for the characteristic function $u_j = j \delta_u,\,j=1,2,3,\dots,M$, the error vector will be given by (here and in the following, $\bm{\theta}$ is suppressed
for simplicity of notation)
\begin{equation}
    \bm{\varepsilon}_N = \left[ \varepsilon_N(u_1), \varepsilon_N(u_2),\dots, \varepsilon_N(u_M) \right]^T.
\end{equation}
It is known that $\bm{\varepsilon_N}$ is asymptotically normal with zero mean and an $(M \times M)$ covariance matrix $\Omega$ \cite{feuerverger-1977,feuerverger-1981},
\begin{equation}
    \Omega_{kl} = \left< \varepsilon_N(u_k) \varepsilon_N(u_l) \right> = C\left(u_k+u_l\right) - C\left(u_k\right)\,C\left( u_l \right).
\end{equation}
An estimator $\widehat{\bm{\theta}}$ of $\bm{\theta}$ can be found by minimizing \cite{tran-1998}
\begin{equation}\label{eq:ecf-est-param-minfun}
    \bm{\varepsilon}_N \widehat{\Omega}^{-1} \bm{\varepsilon}_N^T,
\end{equation}
where $\widehat{\Omega}$ is an estimate of $\Omega$. We estimate the covariance matrix by defining an $(M \times N)$-matrix $E$ with
\begin{equation}\label{eq:ecf-est-param-E}
    E_{kl} = \exp\left(i u_k Y_l\right) - C_N\left(u_k\right).
\end{equation}
Then we have
\begin{equation}\label{eq:ecf-est-param-omega}
    \widehat{\Omega} = \frac{1}{1-N} E E^T.
\end{equation}
Thus, in order to estimate the parameters $\bm{\theta}$, we choose the set of $u$ (as described below) and calculate $\widehat{\Omega}$. With an initial guess $\bm{\theta}_0$, $\bm{\varepsilon}_N$ can be constructed and \Eqref{eq:ecf-est-param-minfun} can be iteratively minimized to find the estimator $\widehat{\bm{\theta}}$.

Choosing the bin size for a histogram can have large effects on the resulting PDF. In the same way, choosing the sampling points $u_j$ has a large effect on the estimation procedure. Various approaches are discussed in \Refs{tran-1998,zhang-2016}. Not many points are needed; both agree on around 10 points as sufficient. Choosing $\delta_u$ is harder and requires the derivative of $C$ with respect to $\bm{\theta}$ (and preferably, the PDF of $Y$, which we do not have). As a low-complexity, high-cost brute force method, we note that $\widehat{\bm{\theta}}$ is insensitive to the initial values for a good choice of $\delta_u$. Thus, we do the estimation for a large range of $\delta_u$ and many different initial values. The results are chosen where many initial values lead to the same result.

\subsection{Examples of parameter estimation}
In this section, we attempt to estimate the parameters $\bm{\theta} = (\gamma,\beta,\epsilon)$ from realizations of the process introduced in \Secref{sec:cf-most-general-fpp}. In the absence of noise, the method works well as long as the degree of pulse overlap is not excessive ($\gamma <20$). We therefore present particularly challenging examples from the full 3-parameter model. Three sets of parameters have been chosen. To describe an FPP with experimentally relevant intermittency level, exponentially distributed amplitudes and low noise level, the parameters $(\gamma,\beta,\epsilon) = (2,0,10^{-1})$ were used. Both pulse overlap and high noise level lead to a distribution more closely resembling a normal distribution. Therefore, in order to reveal the sensitivity to intermittency in a process with high noise level, the parameters $(\gamma,\beta,\epsilon) = (5,1/2,1)$ were used. Lastly, as strong intermittency leads to a more symmetric distribution, revealing the presence of moderate asymmetry in a process with high intermittency was tested with the parameters $(\gamma, \beta, \epsilon) = (10,1/4,10^{-1})$.

\begin{figure}
  \centering
  \subfloat{
    \includegraphics[width = 0.322\textwidth]{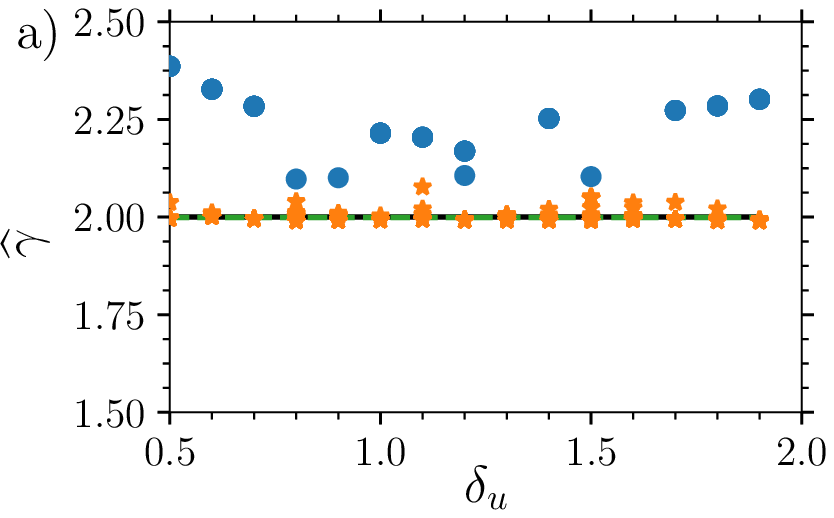}
\label{fig:estECF-g2b0e0.1-G}}
  ~
  \subfloat{
    \includegraphics[width = 0.322\textwidth]{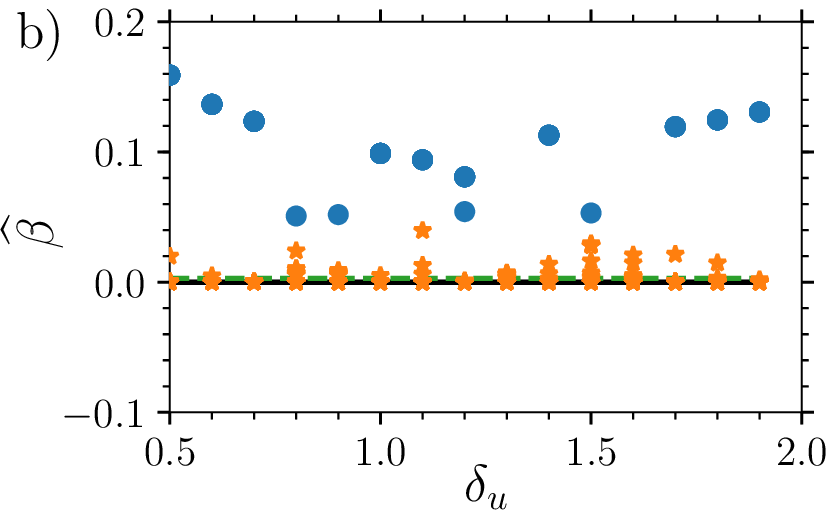}
    \label{fig:estECF-g2b0e0.1-B}}
  ~
  \subfloat{
    \includegraphics[width = 0.322\textwidth]{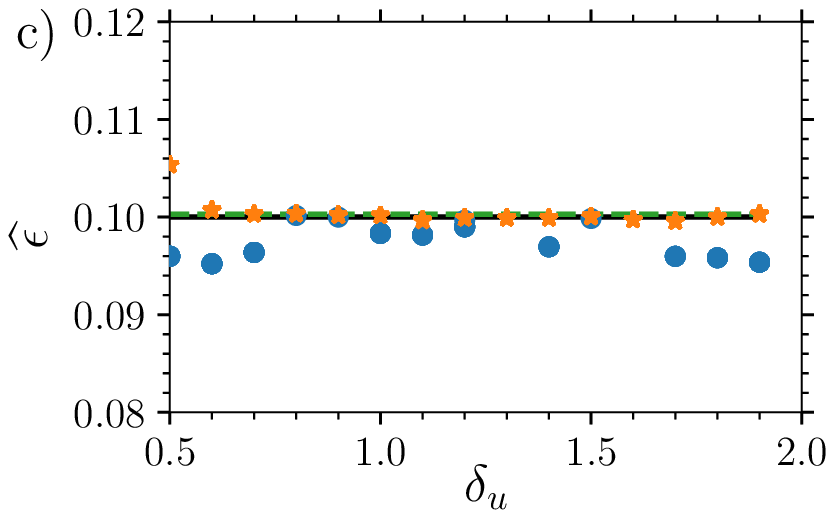}
\label{fig:estECF-g2b0e0.1-E}}

\subfloat{\includegraphics[width = 0.322\textwidth]{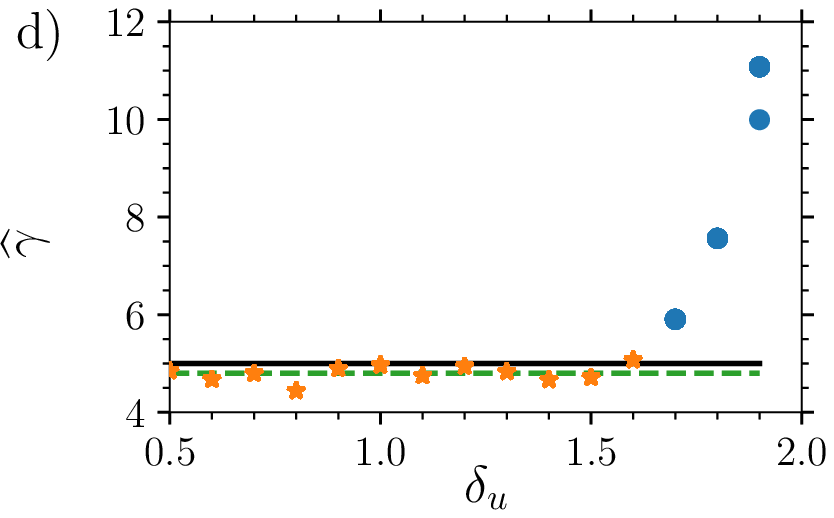}
\label{fig:estECF-g5b0.5e1-G}}
  ~
  \subfloat{
    \includegraphics[width = 0.322\textwidth]{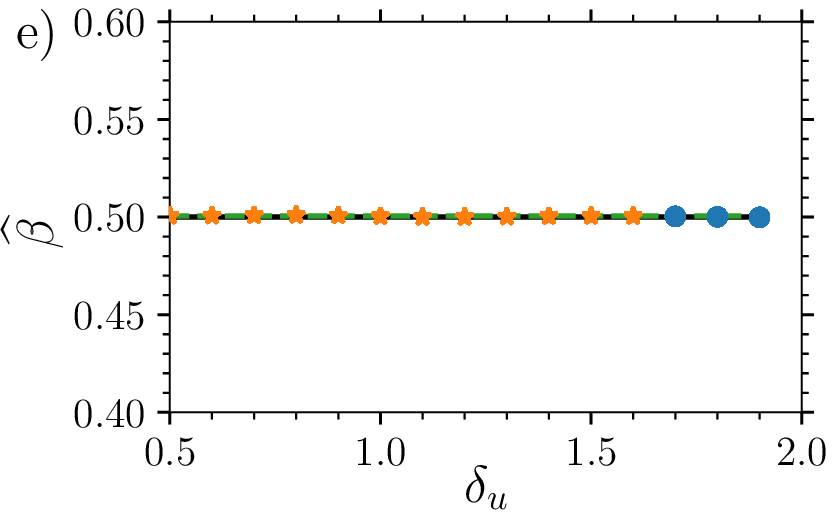}
\label{fig:estECF-g5b0.5e1-B}}
  ~
  \subfloat{
    \includegraphics[width = 0.322\textwidth]{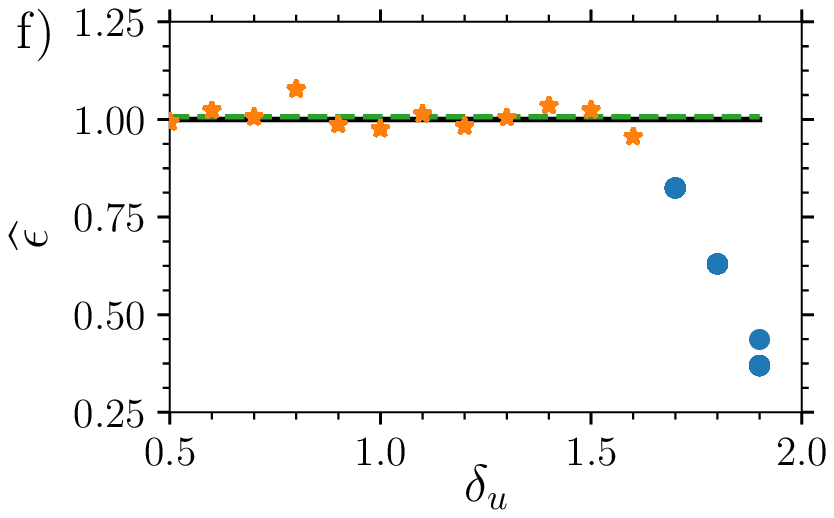}
\label{fig:estECF-g5b0.5e1-E}}
  
  \subfloat{
    \includegraphics[width = 0.322\textwidth]{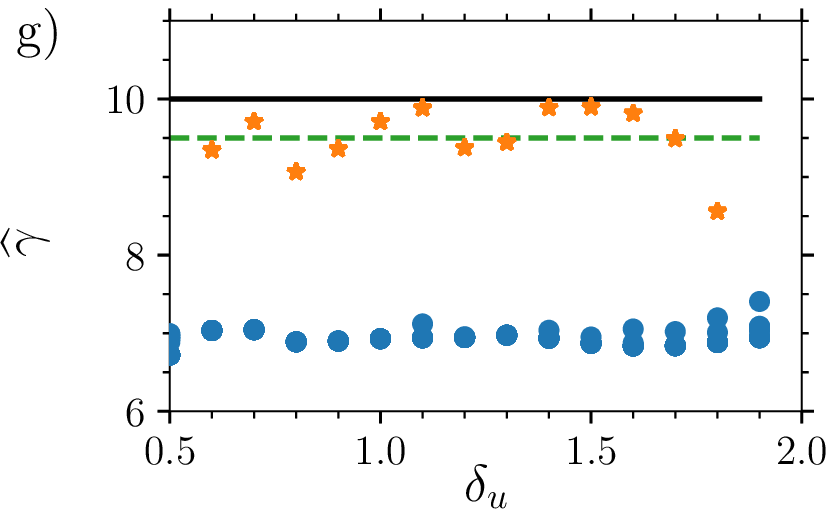}
\label{fig:estECF-g10b0.25e0.1-G}}
  ~
  \subfloat{
    \includegraphics[width = 0.322\textwidth]{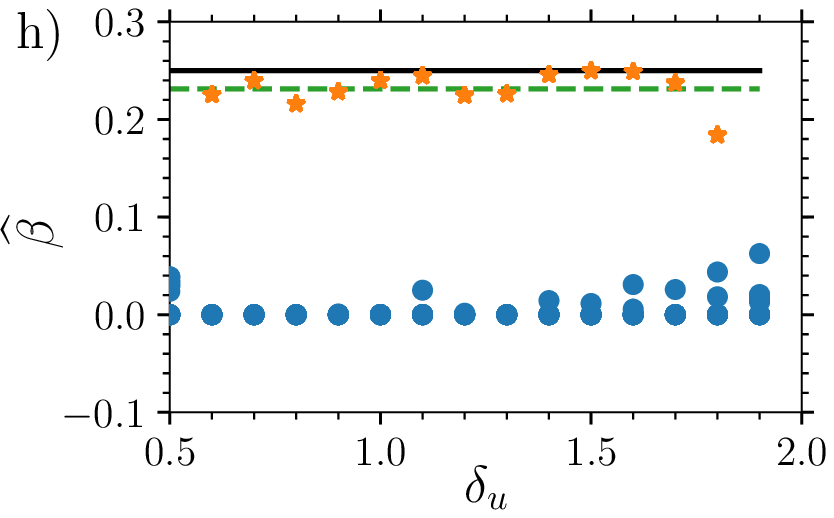}
\label{fig:estECF-g10b0.25e0.1-B}}
  ~
  \subfloat{
    \includegraphics[width = 0.322\textwidth]{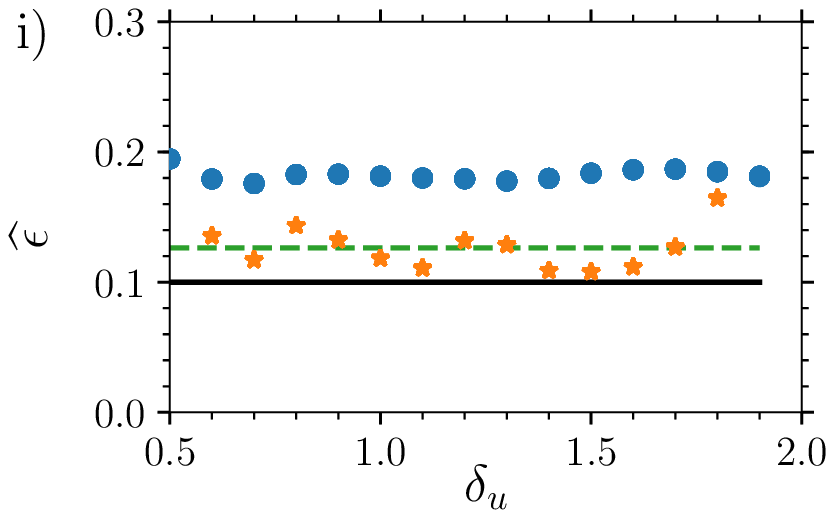}
\label{fig:estECF-g10b0.25e0.1-E}}
  \caption{\label{fig:estECF} Result of parameter estimation procedure. True parameters, given by full black lines, are $(\gamma,\beta,\epsilon) = (2,0,10^{-1})$ (top, a, b, c), $(\gamma,\beta,\epsilon) = (5,1/2,1)$ (middle, d, e, f) and $(\gamma,\beta,\epsilon) = (10,1/4,10^{-1})$ (bottom, g, h, i), with estimated parameters $\widehat{\gamma}$ (left column, a, d, g), $\widehat{\beta}$ (center column, b, e, h) and $\widehat{\epsilon}$ (right column, c, f, i).  For each characteristic value step size $\delta_u$, parameters have been estimated for 18 different sets of initial values, each giving a data point which is either a blue dot or an orange star. Green dashed lines give the mean of values signified by orange stars.}
\end{figure}

In \Figref{fig:estECF}, parameter estimation has been performed for parameters $(\gamma,\beta,\epsilon) = (2,0,10^{-1})$ (top), $(\gamma,\beta,\epsilon) = (5,1/2,1)$ (middle) and $(\gamma,\beta,\epsilon) = (10,1/4,10^{-1})$ (bottom), using the L-BFGS-B - algorithm wrapped by the scipy.optimize.minimize - package \cite{byrd-1995,zhu-1997,scipy}. The synthetic time series have $N=10^6$ data points, and we use $u_j = j \delta_u , \, j = 1,2,\dots,10$ and $\delta_u \in \{0,1,2,\dots,14\}\cdot 10^{-1} + 1/2$. For each $\delta_u$, parameters have been estimated for all combinations of the initial values $\gamma_0 \in \{ 10^{-1},1,10 \}$, $\beta_0 \in \{ 0,1/4,1/2 \}$ and $\epsilon_0 \in \{ 10^{-3},10^{-1} \}$, giving 18 estimated values for each $\delta_u$ and each of $\gamma$, $\beta$ and $\epsilon$. These estimated values are shown by the blue dots and orange stars in \Figref{fig:estECF}. Large scatter for a specific $\delta_u$ signifies high sensitivity to initial values in the estimation procedure. For $\delta_u<0.5$, there was high dependence on initial values in all cases, so we only show results for $\delta_u > 0.5$. In all cases, the black line gives the true value and the green dashed line gives the estimated value. The estimated value is found by taking the mean of all orange stars, while ignoring blue dots. The criteria for deciding which values are used in the estimate are as follows. Top, we assume $\beta = 0$ is known or suspected. Many points cluster around the $\beta=0$ line, and these all correspond to the same value of $\gamma$. Thus, these points are chosen for the parameter estimation, using $\beta<5\times 10^{-2}$ for the orange stars. Middle, the estimates diverge for $\gamma$ and $\epsilon$ in the case $u>1.6$, so these values are not used, and marked with blue dots. Bottom, we have two fixed points for different initial values, one with high $\gamma$ and $\beta$, and low $\epsilon$, and one with low $\gamma$ and $\beta$, and high $\epsilon$. In this case, we choose to explain as little of the variation in the signal as possible with the additional noise level described by $\epsilon$. Low $\epsilon$ corresponds to high $\gamma$ and $\beta$, so $\beta > 10^{-1}$ was used as a condition to mark estimated values with orange stars.

\begin{table}
  \centering
\begin{tabular}{c c c | c | c | c}
  $\gamma$, & $\beta$, & $\epsilon$ & $\widehat{\gamma}-\gamma$ & $\widehat{\beta}-\beta$ & $\widehat{\epsilon}-\epsilon$ \\
  \hline
  $2$, & $0$, & $10^{-1}$ & $0.00 \pm 0.01$ & $0.003 \pm 0.007$ & $0.000 \pm 0.001$ \\
  $5$, & $1/2$, & $1$ & $-0.2 \pm 0.2$ & $0.001 \pm 0.003$ & $0.01 \pm 0.03$ \\
  $10$, & $1/4$, & $10^{-1}$ & $-0.5 \pm 0.4$ & $-0.02 \pm 0.02$ & $0.03 \pm 0.02$
\end{tabular}
  \caption{\label{tab:estECF-params}Table of estimated model parameters with standard deviation for all three sets of example parameters.}
\end{table}

In Table \ref{tab:estECF-params}, the estimated values are presented. This is the true parameter value subtracted from the mean value of the accepted estimated parameters with uncertainty equal to one standard deviation. For the two lowest example intermittency parameters, all estimated parameters are very close to the true value. For the example parameters $(\gamma,\beta,\epsilon) = (10,1/4,10^{-1})$, the procedure underestimates $\gamma$ and overestimates $\epsilon$. This is most likely due to the fact that as $\gamma$ increases, the FPP approaches a normal distribution, making it difficult to separate the FPP from the normally distributed noise in the PDF. Still, all estimated parameters are within two standard deviations of the true parameter values.

\section{Discussion and conclusions}
In this contribution, we have investigated the filtered Poisson process given three different distributions of the pulse amplitudes; exponentially distributed amplitudes, Gamma distributed amplitudes and asymmetrically Laplace distributed amplitudes. For all of these, the mean, variance, and skewness and flatness moments of the resulting process were presented, as well as the parabolic relationship between the skewness and flatness moments. In addition, it was discussed how normally distributed noise affects the moments and the parabolic relation. In all cases, the characteristic function of the filtered Poisson process has a closed form solution, while the probability density function only has closed form solutions for exponentially distributed amplitudes, Gamma distributed amplitudes with shape parameter $\beta = 2$ and symmetrically Laplace distributed amplitudes. 

It was furthermore shown that exponentially distributed amplitudes and Gamma distributed amplitudes with shape parameter $\beta = 2$ lead to PDFs for the FPP which are practically indistinguishable. In previous work \cite{garcia-pop-2017-2}, it was shown that the amplitude distribution does not influence the power spectrum or the auto-correlation function of the FPP. Thus, a realization of an FPP with Gamma distributed amplitudes with shape parameter $\beta = 2$ cannot be easily distinguished from a realization of a FFP with exponentially distributed amplitudes and a slightly larger intermittency parameter $\gamma$. The assumption of Gamma distributed amplitudes therefore leads to more complicated derivations and expressions, but practically equivalent predictions, for the PDF, auto-correlation and power spectrum, and can safely be simplified to the standard assumption of exponentially distributed amplitudes.

Laplace distributed amplitudes lead to expressions which are qualitatively different from exponentially or Gamma distributed amplitudes, since the two latter do not admit negative function values. Only for the case of symmetrically Laplace distributed amplitudes does the process have a closed form PDF. This makes parameter estimation methods requiring the PDF not applicable in general. However, the characteristic function has a closed form expression for any value of the asymmetry parameter. It has been demonstrated that a method based on the empirical characteristic function can be used to reliably estimate the model parameters in realizations of the process. This method can also handle additional noise to the process. The only problem in applying this method is deciding on the sampling points for computing the empirical characteristic function. In this contribution we have disregarded complex, iterative procedures for a simple brute-force method relying on the fact that the estimation procedure should be insensitive to initial parameter guesses for a good choice of sampling points. This method was capable of finding the correct noise ratio and asymmetry parameters in situations complicated by low intermittency (leading to more symmetric and Gaussian-like distributions) in the signal.

The FPP is a reference model for intermittent fluctuations in physical systems, where large amplitude bursts of similar shape dominate the fluctuations. Due to the Poisson process driving the model, it only considers turbulence which is statistically stationary in time. In magnetized plasmas, the model has been successfully applied to measurements of SOL fluctuations, where it is used to systematize and unify measurements. Finding the correct assumptions for the amplitude distributions and having good methods for estimating the model parameters is vital in being able to compare and contrast data for varying plasma parameters and machine configurations. In the future, comparisons between different machines will also be carried out using the framework presented here.

\section*{Acknowledgements}
This work was supported with financial subvention from the Research Council of Norway under grant 240510/F20. The authors acknowledge the generous hospitality of the MIT Plasma Science and Fusion Center where this work was conducted. 

\appendix

\section{Derivation of the characteristic function}\label{app:char-fun-deriv}
In general, it has been assumed that the pulse amplitudes are independently and identically distributed, that the pulse arrival times are independently uniformly distributed, and that the pulses have a fixed shape. The characteristic function of the FPP (and its cumulants) has in this case been discussed in \Refs{rice-1944,rice-1945,parzen-sp,pecseli-fps,lowen-teich-fractal}.

In this section, we will derive the characteristic function of the FPP in a form as general as possible, and investigate exactly which assumptions are necessary in order to obtain a closed form expression. 
In its most general form, the FPP is given by
\begin{equation}
    \Sn_K(t) = \sum\limits_{k=1}^{K(T)} A_k \Snw\left( \frac{t-s_k}{\tau_k},\lambda_k \right),
  \label{app:sn-general}
\end{equation}
where the amplitudes $A$, arrival times $s$, duration times $\tau$ and asymmetries $\lambda$ all are random variables. We begin by assuming that all random variables are independent and identically distributed across pulses, that is for all $k \neq l$:
\begin{equation}
  p_{A_k,\lambda_k,\tau_k,s_k}\left( A_k,\lambda_k,\tau_k,s_k \right) = p_{A_l,\lambda_l,\tau_l,s_l}\left( A_l,\lambda_l,\tau_l,s_l \right),
  \label{app:assumption-1}
\end{equation}
and
\begin{multline}
p_{A_k,\lambda_k,\tau_k,s_k,A_l,\lambda_l,\tau_l,s_l}\left( A_k,\lambda_k,\tau_k,s_k,  A_l,\lambda_l,\tau_l,s_l \right) = \\ p_{A_k,\lambda_k,\tau_k,s_k}\left( A_k,\lambda_k,\tau_k,s_k \right)p_{A_l,\lambda_l,\tau_l,s_l}\left( A_l,\lambda_l,\tau_l,s_l \right).
  \label{app:assumption-2}
\end{multline}
One could easily imagine an alternative FPP where for example the amplitude of the pulse depends on the waiting time from the last pulse. In this case, $A_k$ would depend on $s_k$ and $s_{k-1}$, and a very different treatment from the present one would be required.
In the following we drop the index on the random variables, since each pulse is statistically identical. The characteristic function of $\Sn$ is the product of all characteristic functions of $\phi = A \Snw\left( \frac{t-s}{\tau},\lambda \right)$. Fixing $K$ for the moment, we have
\begin{equation}
  C_\Sn(u;K,t) = \prod\limits_{k=1}^K C_{\phi}(u;t)=C_{\phi}(u;t)^K,
  \label{app:char-fun-prod-def}
\end{equation}
where the variables after the semicolon are parameters in the characteristic function. By definition,
\begin{equation}
  C_{\phi}(u;t) = \left< \exp\left(i u \phi\right) \right> = \int\limits_{-\infty}^\infty \rmd A\,\int\limits_{-\infty}^\infty \rmd \lambda\,\int\limits_{-\infty}^\infty \rmd \tau\,\int\limits_{-\infty}^\infty \rmd s\, p_{A,\lambda,\tau,s}(A,\lambda,\tau,s) \exp\left( i A u \Snw\left( \frac{t-s}{\tau},\lambda \right) \right).
  \label{eq:char-fun-snw-def}
\end{equation}

The PDF of $\Sn$ for fixed $K$ is
\begin{equation}
  P_\Sn(\Sn|K) = \frac{1}{2 \pi} \int\limits_{-\infty}^\infty \rmd u\, \exp\left( i u \Sn \right) C_{\phi}(u;t)^K.
  \label{app:pdf-sn-def}
\end{equation}
Using that $K$ is Poisson distributed,
\begin{equation}
  P_K(K|T) = \frac{1}{K!} \left( \frac{T}{\tw} \right)^K \exp\left( -\frac{T}{\tw} \right),
  \label{app:K-poisson}
\end{equation}
we have 
\begin{equation}
  P_\Sn(\Sn|T) = \sum_{K=0}^\infty P_\Sn(\Sn|K)P_K(K|T) = \frac{1}{2 \pi} \int\limits_{-\infty}^\infty \rmd u\, \exp\left( i u \Sn \right) \exp\left\{ \frac{T}{\tw} \left[ C_{\phi}(u;t)-1\right]\right\}.
  \label{app:pdf-sn}
\end{equation}
The expression inside the last exponential function can be identified as the logarithm of $C_\Sn(u)$. Since the joint PDF of $A$, $\lambda$, $\tau$ and $s$ integrates to $1$ by definition, we have that this expression is
\begin{equation}
  \ln C_\Sn(u;T,t) = \frac{T}{\tw} \int\limits_{-\infty}^\infty \rmd A\,\int\limits_{-\infty}^\infty \rmd \lambda\,\int\limits_{-\infty}^\infty \rmd \tau\,\int\limits_{-\infty}^\infty \rmd s\, p_{A,\lambda,\tau,s}(A,\lambda,\tau,s) \left[ \exp\left( i A u \Snw\left( \frac{t-s}{\tau},\lambda \right) \right) -1 \right],
  \label{app:char-fun-sn-der1}
\end{equation}
Since $K$ is Poisson distributed, $s$ is uniformly distributed on $0 \leq s \leq T$. Assuming that $s$ is independent of all the other random variables, we have
\begin{equation}
  \ln C_\Sn(u;T,t) = \frac{1}{\tw} \int\limits_{-\infty}^\infty \rmd A\,\int\limits_{-\infty}^\infty \rmd \lambda\,\int\limits_{-\infty}^\infty \rmd \tau\, p_{A,\lambda,\tau}(A,\lambda,\tau) \int\limits_{0}^T \rmd s\, \left[ \exp\left( i A u \Snw\left( \frac{t-s}{\tau},\lambda \right) \right) -1 \right].
  \label{app:char-fun-sn-der2}
\end{equation}
Exchanging the variable $s$ to $\theta = (t-s)/\tau$, assuming stationarity and ignoring end effects for $\theta$ (that is, setting the integration limits back to $-\infty < \theta <\infty$), we have
\begin{equation}
  \ln C_\Sn(u) = \frac{1}{\tw} \int\limits_{-\infty}^\infty \rmd A\,\int\limits_{-\infty}^\infty \rmd \lambda\,\int\limits_{-\infty}^\infty \rmd \tau\, \tau p_{A,\lambda,\tau}(A,\lambda,\tau) \int\limits_{-\infty}^\infty \rmd \theta\, \left[ \exp\left( i A u \Snw\left( \theta ,\lambda \right) \right) -1 \right].
  \label{app:char-fun-sn-der3}
\end{equation}
This is the most general form of the characteristic function of $\Sn$, where we only assume that pulses are independent of each other and that the arrivals follow a Poisson process and are independent of the other properties of the pulses.

Assuming that $A$, $\lambda$ and $\tau$ are independent, and that $\lambda$ takes on a specific value, we arrive at
\begin{equation}
  \ln C_\Sn(u) = \gamma \int\limits_{-\infty}^\infty \rmd A\, P_A(A) \int\limits_{-\infty}^\infty \rmd \theta\, \left[ \exp\left( i u A \Snw(\theta) \right)-1 \right].
  \label{app:char-fun-sn-integral-start}
\end{equation}
Changing the order of integration in \Eqref{app:char-fun-sn-integral-start} and using the definition of the characteristic function, we have \cite{garcia-pop-2017-2}
\begin{equation}
  \ln C_\Sn(u) = \gamma \int\limits_{-\infty}^\infty \rmd \theta \,\left[ -1 + \int\limits_{-\infty}^\infty \rmd A\, P_A(A) \exp\left( i u A \Snw(\theta) \right) \right] = \gamma \int\limits_{-\infty}^\infty \rmd \theta\, \left[ C_A\left( u\, \Snw(\theta) \right)-1 \right],
  \label{app:char-fun-sn-integral}
\end{equation}
where $C_A$ is the characteristic function for the amplitude distribution $P_A$. 

If we instead expand the exponential function in \Eqref{app:char-fun-sn-der3} into a sum, we have
\begin{equation}
  \ln C_\Sn(u;T,t) =\frac{1}{\tw} \sum_{n=1}^\infty \frac{(i u)^n}{n!} \int\limits_{-\infty}^\infty \rmd A\,\int\limits_{-\infty}^\infty \rmd \lambda\,\int\limits_{-\infty}^\infty \rmd \tau\, \tau p_{A,\lambda,\tau}(A,\lambda,\tau) A^n \int\limits_{-\infty}^\infty \rmd \theta\,\Snw\left( \theta,\lambda \right)^n.
  \label{app:char-fun-sn-der4}
\end{equation}
The last integral gives $I_n(\lambda)$, and
\begin{equation}
  \ln C_\Sn(u) = \frac{1}{\tw} \sum_{n=1}^\infty \frac{(i u)^n}{n!} \left< \tau A^n I_n(\lambda) \right>.
  \label{app:char-fun-general}
\end{equation}
For a stationary FPP with independent pulses, where the pulse does not depend on its arrival or waiting time, this is the most general form of the characteristic function, as it allows for any relationship between the variables determining the pulse shape (amplitude, decay time and asymmetry). It also reveals the cumulants,
\begin{equation}
  \kappa_n = \frac{1}{\tw} \mean{\tau A^n I_n(\lambda)}.
  \label{app:char-fun-cumulants}
\end{equation}
If we use the two-sided exponential pulse, $I_n$ is independent of $\lambda$, and thus $\lambda$ (and its dependence on $A$ and $\tau$) plays no further role in \Eqref{app:char-fun-general}. If we additionally assume that $\tau$ and $A$ are independent, we have 
\begin{equation}
  \ln C_\Sn(u) = \gamma \sum_{n=1}^\infty \frac{(i u)^n}{n!} \left< A^n \right> I_n,
  \label{app:char-fun-sn-end}
\end{equation}
where $\gamma = \td/\tw = \left< \tau \right> / \tw$. Crucially, this expression (and thus the PDF and moments of $\Sn$) is independent of the distributions of $\tau$ and $\lambda$ as long as $\tau$ and $A$ are independent and the pulse shape has $I_n$ independent of $\lambda$.  

\section{The distribution of a sum of independent random variables}\label{app:sum-rv}
For simplicity, we include some well known properties of the sum of characteristic functions, see for example \cite{stark-psrpe}. Given two independent random variables, $Y_1$ and $Y_2$ with respective characteristic functions $C_{Y_1}$ and $C_{Y_2}$ and respective cumulants $\kappa_n^{Y_1}$ and $\kappa_n^{Y_2}$, their sum
\begin{equation}
  Y = Y_1 + Y_2
  \label{app:sum-rv}
\end{equation}
has the characteristic function
\begin{equation}
  C_Y(u) = C_{Y_1}(u) C_{Y_2}(u)
  \label{app:sum-rv-cf}
\end{equation}
and the probability density function
\begin{equation}
  P_Y(Y) = \left[ P_{Y_1} * P_{Y_2} \right](Y),
  \label{app:sum-rv-pdf}
\end{equation}
where $*$ denotes convolution. The cumulants $\kappa_n^Y$ of $Y$ are found by
\begin{align}
  \ln C_Y(u) &= \ln\left[ C_{Y_1}(u) \, C_{Y_2}(u) \right] \nonumber\\
  &= \ln C_{Y_1}(u) + \ln C_{Y_2}(u) \nonumber \\
  \sum\limits_{n=1}^{\infty} \kappa^Y_n \frac{(i u)^n}{n!} &= \sum\limits_{n=1}^{\infty} \kappa^{Y_1}_n \frac{(i u)^n}{n!} + \sum\limits_{n=1}^{\infty} \kappa^{Y_2}_n \frac{(i u)^n}{n!} \nonumber \\
  \kappa^Y_n &= \kappa^{Y_1}_n + \kappa^{Y_2}_n.
  \label{app:sum-rv-cumulants}
\end{align}

\section{PDF of filtered Poisson process with gamma distributed amplitudes}\label{app:sn-gamma-amp-cf-to-pdf}
In \Eqref{eq:cf-k2-sn-gamma-amp}, the characteristic function of the filtered Poisson process with Gamma distributed amplitudes with shape parameter $\beta=2$ is shown to be
\begin{equation}
  C_\Sn(u) = \exp(-\gamma) \exp\left(\frac{\gamma}{1-i \alpha u} \right) (1-i \alpha u)^{-\gamma}.
  \label{app:cf-k2-sn-gamma-amp}
\end{equation}
This characteristic function is the product of the two functions
\begin{equation}
  g_1(u) = \exp(-\gamma)\exp\left(\frac{\gamma}{1-i \alpha u} \right)
  \label{app:cf-sn-gamma-amp-part1}
\end{equation}
and
\begin{equation}
  g_2(u) = (1-i \alpha u)^{-\gamma}.
  \label{app:cf-sn-gamma-amp-part2}
\end{equation}
The PDF of $\Sn$ is thus the convolution of the inverse Fourier transforms of these functions, $P_\Sn(\Sn) = G_1(\Sn)*G_2(\Sn)$, where $G_1$ is the inverse Fourier transform of $g_1$, and $G_2$ is the inverse Fourier transform of $g_2$. The second function $g_2$ is the characteristic function of a Gamma distributed variable with shape parameter $\gamma$ and scale parameter $\alpha$, so
\begin{equation}
  G_2(\Sn) = \frac{\Sn^{\gamma-1}}{\alpha^\gamma \Gamma(\gamma)} \exp(-\Sn/\alpha),\,\Sn>0.
  \label{app:inv-trans-part2}
\end{equation}
The inverse transform of the first part is easy to see if we expand the exponential function into a sum,
\begin{equation}
  g_1(u) = \exp(-\gamma)\sum\limits_{n=0}^\infty \frac{\gamma^n}{n! (1-i \alpha u)^n}.
  \label{app:cf-sn-gamma-amp-part1-alt}
\end{equation}
For $n=0$, the inverse transform is a Dirac delta function. For $n>0$, this is some factor multiplied by the characteristic function of a Gamma distributed random variable with shape parameter $n$ and scale parameter $\alpha$,
\begin{equation}
  G_1(\Sn) = \exp(-\gamma) \left[ \delta(\Sn) + \sum\limits_{n=1}^\infty \frac{\gamma^n}{n!} \frac{\Sn^{n-1}}{\alpha^n \Gamma(n)}\exp(-\Sn/\alpha) \right],\, \Sn>0.
  \label{app:inv-trans-part1}
\end{equation}
Performing the convolution gives
\begin{align}
  P_\Sn(\Sn) &= \exp(-\gamma) \left[ \frac{\Sn^{\gamma-1}}{\alpha^\gamma \Gamma(\gamma)} \exp\left(-\frac{\Sn}{\alpha}\right) + \sum\limits_{n=1}^{\infty} \frac{\gamma^n}{n!} \exp\left(-\frac{\Sn}{\alpha}\right) \frac{\Sn^{\gamma+n-1}}{\alpha^{\gamma+n} \Gamma(\gamma+n)}\right] \nonumber\\
  &= \frac{1}{\alpha} \exp\left(-\gamma-\frac{\Sn}{\alpha}\right) \sum_{n=0}^\infty \frac{\gamma^n}{n!} \left(\frac{\Sn}{\alpha}\right)^{\gamma+n-1}\frac{1}{\Gamma(\gamma+n)} \nonumber\\
  &= \frac{1}{\alpha} \left( \frac{\Sn}{\gamma \alpha} \right)^{(\gamma-1)/2} \exp\left(-\gamma-\frac{\Sn}{\alpha}\right)  I_{\gamma-1} \left( 2 \sqrt{\gamma \frac{\Sn}{\alpha}} \right),
  \label{app:pdf-sn-gamma-amp}
\end{align}
where $I$ is the modified Bessel function of the first kind \cite{nist-dlmf}.

%


\begin{thebibliography}{64}%
\makeatletter
\providecommand \@ifxundefined [1]{%
 \@ifx{#1\undefined}
}%
\providecommand \@ifnum [1]{%
 \ifnum #1\expandafter \@firstoftwo
 \else \expandafter \@secondoftwo
 \fi
}%
\providecommand \@ifx [1]{%
 \ifx #1\expandafter \@firstoftwo
 \else \expandafter \@secondoftwo
 \fi
}%
\providecommand \natexlab [1]{#1}%
\providecommand \enquote  [1]{``#1''}%
\providecommand \bibnamefont  [1]{#1}%
\providecommand \bibfnamefont [1]{#1}%
\providecommand \citenamefont [1]{#1}%
\providecommand \href@noop [0]{\@secondoftwo}%
\providecommand \href [0]{\begingroup \@sanitize@url \@href}%
\providecommand \@href[1]{\@@startlink{#1}\@@href}%
\providecommand \@@href[1]{\endgroup#1\@@endlink}%
\providecommand \@sanitize@url [0]{\catcode `\\12\catcode `\$12\catcode
  `\&12\catcode `\#12\catcode `\^12\catcode `\_12\catcode `\%12\relax}%
\providecommand \@@startlink[1]{}%
\providecommand \@@endlink[0]{}%
\providecommand \url  [0]{\begingroup\@sanitize@url \@url }%
\providecommand \@url [1]{\endgroup\@href {#1}{\urlprefix }}%
\providecommand \urlprefix  [0]{URL }%
\providecommand \Eprint [0]{\href }%
\providecommand \doibase [0]{http://dx.doi.org/}%
\providecommand \selectlanguage [0]{\@gobble}%
\providecommand \bibinfo  [0]{\@secondoftwo}%
\providecommand \bibfield  [0]{\@secondoftwo}%
\providecommand \translation [1]{[#1]}%
\providecommand \BibitemOpen [0]{}%
\providecommand \bibitemStop [0]{}%
\providecommand \bibitemNoStop [0]{.\EOS\space}%
\providecommand \EOS [0]{\spacefactor3000\relax}%
\providecommand \BibitemShut  [1]{\csname bibitem#1\endcsname}%
\let\auto@bib@innerbib\@empty
\bibitem [{\citenamefont {LaBombard}\ \emph {et~al.}(2001)\citenamefont
  {LaBombard}, \citenamefont {Boivin}, \citenamefont {Greenwald}, \citenamefont
  {Hughes}, \citenamefont {Lipschultz}, \citenamefont {Mossessian},
  \citenamefont {Pitcher}, \citenamefont {Terry}, \citenamefont {Zweben},\ and\
  \citenamefont {{Alcator Group}}}]{labombard-2001}%
  \BibitemOpen
  \bibfield  {author} {\bibinfo {author} {\bibfnamefont {B.}~\bibnamefont
  {LaBombard}}, \bibinfo {author} {\bibfnamefont {R.~L.}\ \bibnamefont
  {Boivin}}, \bibinfo {author} {\bibfnamefont {M.}~\bibnamefont {Greenwald}},
  \bibinfo {author} {\bibfnamefont {J.}~\bibnamefont {Hughes}}, \bibinfo
  {author} {\bibfnamefont {B.}~\bibnamefont {Lipschultz}}, \bibinfo {author}
  {\bibfnamefont {D.}~\bibnamefont {Mossessian}}, \bibinfo {author}
  {\bibfnamefont {C.~S.}\ \bibnamefont {Pitcher}}, \bibinfo {author}
  {\bibfnamefont {J.~L.}\ \bibnamefont {Terry}}, \bibinfo {author}
  {\bibfnamefont {S.~J.}\ \bibnamefont {Zweben}}, \ and\ \bibinfo {author}
  {\bibnamefont {{Alcator Group}}},\ }\href {\doibase 10.1063/1.1352596}
  {\bibfield  {journal} {\bibinfo  {journal} {Phys. Plasmas}\ }\textbf
  {\bibinfo {volume} {8}},\ \bibinfo {pages} {2107} (\bibinfo {year}
  {2001})}\BibitemShut {NoStop}%
\bibitem [{\citenamefont {Krasheninnikov}(2001)}]{krasheninnikov-2001}%
  \BibitemOpen
  \bibfield  {author} {\bibinfo {author} {\bibfnamefont {S.}~\bibnamefont
  {Krasheninnikov}},\ }\href {\doibase 10.1016/S0375-9601(01)00252-3}
  {\bibfield  {journal} {\bibinfo  {journal} {Phys. Lett. A}\ }\textbf
  {\bibinfo {volume} {283}},\ \bibinfo {pages} {368} (\bibinfo {year}
  {2001})}\BibitemShut {NoStop}%
\bibitem [{\citenamefont {Rudakov}\ \emph {et~al.}(2002)\citenamefont
  {Rudakov}, \citenamefont {Boedo}, \citenamefont {Moyer}, \citenamefont
  {Krasheninnikov}, \citenamefont {Leonard}, \citenamefont {Mahdavi},
  \citenamefont {McKee}, \citenamefont {Porter}, \citenamefont {Stangeby},
  \citenamefont {Watkins}, \citenamefont {West}, \citenamefont {Whyte},\ and\
  \citenamefont {Antar}}]{rudakov-2002}%
  \BibitemOpen
  \bibfield  {author} {\bibinfo {author} {\bibfnamefont {D.~L.}\ \bibnamefont
  {Rudakov}}, \bibinfo {author} {\bibfnamefont {J.~A.}\ \bibnamefont {Boedo}},
  \bibinfo {author} {\bibfnamefont {R.~A.}\ \bibnamefont {Moyer}}, \bibinfo
  {author} {\bibfnamefont {S.}~\bibnamefont {Krasheninnikov}}, \bibinfo
  {author} {\bibfnamefont {A.~W.}\ \bibnamefont {Leonard}}, \bibinfo {author}
  {\bibfnamefont {M.~A.}\ \bibnamefont {Mahdavi}}, \bibinfo {author}
  {\bibfnamefont {G.~R.}\ \bibnamefont {McKee}}, \bibinfo {author}
  {\bibfnamefont {G.~D.}\ \bibnamefont {Porter}}, \bibinfo {author}
  {\bibfnamefont {P.~C.}\ \bibnamefont {Stangeby}}, \bibinfo {author}
  {\bibfnamefont {J.~G.}\ \bibnamefont {Watkins}}, \bibinfo {author}
  {\bibfnamefont {W.~P.}\ \bibnamefont {West}}, \bibinfo {author}
  {\bibfnamefont {D.~G.}\ \bibnamefont {Whyte}}, \ and\ \bibinfo {author}
  {\bibfnamefont {G.}~\bibnamefont {Antar}},\ }\href {\doibase
  10.1088/0741-3335/44/6/308} {\bibfield  {journal} {\bibinfo  {journal}
  {Plasma Phys. Control. Fusion}\ }\textbf {\bibinfo {volume} {44}},\ \bibinfo
  {pages} {308} (\bibinfo {year} {2002})}\BibitemShut {NoStop}%
\bibitem [{\citenamefont {D'Ippolito}\ \emph {et~al.}(2004)\citenamefont
  {D'Ippolito}, \citenamefont {Myra}, \citenamefont {Krasheninnikov},
  \citenamefont {Yu},\ and\ \citenamefont {Pigarov}}]{dippolito-2004}%
  \BibitemOpen
  \bibfield  {author} {\bibinfo {author} {\bibfnamefont {D.~A.}\ \bibnamefont
  {D'Ippolito}}, \bibinfo {author} {\bibfnamefont {J.~R.}\ \bibnamefont
  {Myra}}, \bibinfo {author} {\bibfnamefont {S.~I.}\ \bibnamefont
  {Krasheninnikov}}, \bibinfo {author} {\bibfnamefont {G.~Q.}\ \bibnamefont
  {Yu}}, \ and\ \bibinfo {author} {\bibfnamefont {A.~Y.}\ \bibnamefont
  {Pigarov}},\ }\href {\doibase 10.1002/ctpp.200410030} {\bibfield  {journal}
  {\bibinfo  {journal} {Contrib. to Plasma Phys.}\ }\textbf {\bibinfo {volume}
  {44}},\ \bibinfo {pages} {205} (\bibinfo {year} {2004})}\BibitemShut
  {NoStop}%
\bibitem [{\citenamefont {Rudakov}\ \emph {et~al.}(2005)\citenamefont
  {Rudakov}, \citenamefont {Boedo}, \citenamefont {Moyer}, \citenamefont
  {Stangeby}, \citenamefont {Watkins}, \citenamefont {Whyte}, \citenamefont
  {Zeng}, \citenamefont {Brooks}, \citenamefont {Doerner}, \citenamefont
  {Evans}, \citenamefont {Fenstermacher}, \citenamefont {Groth}, \citenamefont
  {Hollmann}, \citenamefont {Krasheninnikov}, \citenamefont {Lasnier},
  \citenamefont {Leonard}, \citenamefont {Mahdavi}, \citenamefont {McKee},
  \citenamefont {McLean}, \citenamefont {Pigarov}, \citenamefont {Wampler},
  \citenamefont {Wang}, \citenamefont {West},\ and\ \citenamefont
  {Wong}}]{rudakov-2005}%
  \BibitemOpen
  \bibfield  {author} {\bibinfo {author} {\bibfnamefont {D.}~\bibnamefont
  {Rudakov}}, \bibinfo {author} {\bibfnamefont {J.}~\bibnamefont {Boedo}},
  \bibinfo {author} {\bibfnamefont {R.}~\bibnamefont {Moyer}}, \bibinfo
  {author} {\bibfnamefont {P.}~\bibnamefont {Stangeby}}, \bibinfo {author}
  {\bibfnamefont {J.}~\bibnamefont {Watkins}}, \bibinfo {author} {\bibfnamefont
  {D.}~\bibnamefont {Whyte}}, \bibinfo {author} {\bibfnamefont
  {L.}~\bibnamefont {Zeng}}, \bibinfo {author} {\bibfnamefont {N.}~\bibnamefont
  {Brooks}}, \bibinfo {author} {\bibfnamefont {R.}~\bibnamefont {Doerner}},
  \bibinfo {author} {\bibfnamefont {T.}~\bibnamefont {Evans}}, \bibinfo
  {author} {\bibfnamefont {M.}~\bibnamefont {Fenstermacher}}, \bibinfo {author}
  {\bibfnamefont {M.}~\bibnamefont {Groth}}, \bibinfo {author} {\bibfnamefont
  {E.}~\bibnamefont {Hollmann}}, \bibinfo {author} {\bibfnamefont
  {S.}~\bibnamefont {Krasheninnikov}}, \bibinfo {author} {\bibfnamefont
  {C.}~\bibnamefont {Lasnier}}, \bibinfo {author} {\bibfnamefont
  {A.}~\bibnamefont {Leonard}}, \bibinfo {author} {\bibfnamefont
  {M.}~\bibnamefont {Mahdavi}}, \bibinfo {author} {\bibfnamefont
  {G.}~\bibnamefont {McKee}}, \bibinfo {author} {\bibfnamefont
  {A.}~\bibnamefont {McLean}}, \bibinfo {author} {\bibfnamefont
  {A.}~\bibnamefont {Pigarov}}, \bibinfo {author} {\bibfnamefont
  {W.}~\bibnamefont {Wampler}}, \bibinfo {author} {\bibfnamefont
  {G.}~\bibnamefont {Wang}}, \bibinfo {author} {\bibfnamefont {W.}~\bibnamefont
  {West}}, \ and\ \bibinfo {author} {\bibfnamefont {C.}~\bibnamefont {Wong}},\
  }\href {\doibase 10.1088/0029-5515/45/12/014} {\bibfield  {journal} {\bibinfo
   {journal} {Nucl. Fusion}\ }\textbf {\bibinfo {volume} {45}},\ \bibinfo
  {pages} {1589} (\bibinfo {year} {2005})}\BibitemShut {NoStop}%
\bibitem [{\citenamefont {Zweben}\ \emph {et~al.}(2007)\citenamefont {Zweben},
  \citenamefont {Boedo}, \citenamefont {Grulke}, \citenamefont {Hidalgo},
  \citenamefont {LaBombard}, \citenamefont {Maqueda}, \citenamefont {Scarin},\
  and\ \citenamefont {Terry}}]{zweben-2007}%
  \BibitemOpen
  \bibfield  {author} {\bibinfo {author} {\bibfnamefont {S.~J.}\ \bibnamefont
  {Zweben}}, \bibinfo {author} {\bibfnamefont {J.~A.}\ \bibnamefont {Boedo}},
  \bibinfo {author} {\bibfnamefont {O.}~\bibnamefont {Grulke}}, \bibinfo
  {author} {\bibfnamefont {C.}~\bibnamefont {Hidalgo}}, \bibinfo {author}
  {\bibfnamefont {B.}~\bibnamefont {LaBombard}}, \bibinfo {author}
  {\bibfnamefont {R.~J.}\ \bibnamefont {Maqueda}}, \bibinfo {author}
  {\bibfnamefont {P.}~\bibnamefont {Scarin}}, \ and\ \bibinfo {author}
  {\bibfnamefont {J.~L.}\ \bibnamefont {Terry}},\ }\href {\doibase
  10.1088/0741-3335/49/7/S01} {\bibfield  {journal} {\bibinfo  {journal}
  {Plasma Phys. Control. Fusion}\ }\textbf {\bibinfo {volume} {49}},\ \bibinfo
  {pages} {S1} (\bibinfo {year} {2007})}\BibitemShut {NoStop}%
\bibitem [{\citenamefont {Krasheninnikov}\ \emph {et~al.}(2008)\citenamefont
  {Krasheninnikov}, \citenamefont {D'Ippolito},\ and\ \citenamefont
  {Myra}}]{krasheninnikov-2008}%
  \BibitemOpen
  \bibfield  {author} {\bibinfo {author} {\bibfnamefont {S.~I.}\ \bibnamefont
  {Krasheninnikov}}, \bibinfo {author} {\bibfnamefont {D.~A.}\ \bibnamefont
  {D'Ippolito}}, \ and\ \bibinfo {author} {\bibfnamefont {J.~R.}\ \bibnamefont
  {Myra}},\ }\href {\doibase 10.1017/S0022377807006940} {\bibfield  {journal}
  {\bibinfo  {journal} {J. Plasma Phys.}\ }\textbf {\bibinfo {volume} {74}},\
  \bibinfo {pages} {679} (\bibinfo {year} {2008})}\BibitemShut {NoStop}%
\bibitem [{\citenamefont {GARCIA}(2009)}]{garcia-2009}%
  \BibitemOpen
  \bibfield  {author} {\bibinfo {author} {\bibfnamefont {O.~E.}\ \bibnamefont
  {GARCIA}},\ }\href {\doibase 10.1585/pfr.4.019} {\bibfield  {journal}
  {\bibinfo  {journal} {Plasma Fusion Res.}\ }\textbf {\bibinfo {volume} {4}},\
  \bibinfo {pages} {019} (\bibinfo {year} {2009})}\BibitemShut {NoStop}%
\bibitem [{\citenamefont {D'Ippolito}\ \emph {et~al.}(2011)\citenamefont
  {D'Ippolito}, \citenamefont {Myra},\ and\ \citenamefont
  {Zweben}}]{dippolito-2011}%
  \BibitemOpen
  \bibfield  {author} {\bibinfo {author} {\bibfnamefont {D.~A.}\ \bibnamefont
  {D'Ippolito}}, \bibinfo {author} {\bibfnamefont {J.~R.}\ \bibnamefont
  {Myra}}, \ and\ \bibinfo {author} {\bibfnamefont {S.~J.}\ \bibnamefont
  {Zweben}},\ }\href {\doibase 10.1063/1.3594609} {\bibfield  {journal}
  {\bibinfo  {journal} {Phys. Plasmas}\ }\textbf {\bibinfo {volume} {18}},\
  \bibinfo {pages} {060501} (\bibinfo {year} {2011})}\BibitemShut {NoStop}%
\bibitem [{\citenamefont {Kube}\ \emph {et~al.}(2013)\citenamefont {Kube},
  \citenamefont {Garcia}, \citenamefont {LaBombard}, \citenamefont {Terry},\
  and\ \citenamefont {Zweben}}]{kube-2013}%
  \BibitemOpen
  \bibfield  {author} {\bibinfo {author} {\bibfnamefont {R.}~\bibnamefont
  {Kube}}, \bibinfo {author} {\bibfnamefont {O.}~\bibnamefont {Garcia}},
  \bibinfo {author} {\bibfnamefont {B.}~\bibnamefont {LaBombard}}, \bibinfo
  {author} {\bibfnamefont {J.}~\bibnamefont {Terry}}, \ and\ \bibinfo {author}
  {\bibfnamefont {S.}~\bibnamefont {Zweben}},\ }\href {\doibase
  10.1016/j.jnucmat.2013.01.104} {\bibfield  {journal} {\bibinfo  {journal} {J.
  Nucl. Mater.}\ }\textbf {\bibinfo {volume} {438}},\ \bibinfo {pages} {S505}
  (\bibinfo {year} {2013})}\BibitemShut {NoStop}%
\bibitem [{\citenamefont {Zweben}\ \emph {et~al.}(2015)\citenamefont {Zweben},
  \citenamefont {Davis}, \citenamefont {Kaye}, \citenamefont {Myra},
  \citenamefont {Bell}, \citenamefont {LeBlanc}, \citenamefont {Maqueda},
  \citenamefont {Munsat}, \citenamefont {Sabbagh}, \citenamefont {Sechrest},\
  and\ \citenamefont {Stotler}}]{zweben-2015}%
  \BibitemOpen
  \bibfield  {author} {\bibinfo {author} {\bibfnamefont {S.}~\bibnamefont
  {Zweben}}, \bibinfo {author} {\bibfnamefont {W.}~\bibnamefont {Davis}},
  \bibinfo {author} {\bibfnamefont {S.}~\bibnamefont {Kaye}}, \bibinfo {author}
  {\bibfnamefont {J.}~\bibnamefont {Myra}}, \bibinfo {author} {\bibfnamefont
  {R.}~\bibnamefont {Bell}}, \bibinfo {author} {\bibfnamefont {B.}~\bibnamefont
  {LeBlanc}}, \bibinfo {author} {\bibfnamefont {R.}~\bibnamefont {Maqueda}},
  \bibinfo {author} {\bibfnamefont {T.}~\bibnamefont {Munsat}}, \bibinfo
  {author} {\bibfnamefont {S.}~\bibnamefont {Sabbagh}}, \bibinfo {author}
  {\bibfnamefont {Y.}~\bibnamefont {Sechrest}}, \ and\ \bibinfo {author}
  {\bibfnamefont {D.}~\bibnamefont {Stotler}},\ }\href {\doibase
  10.1088/0029-5515/55/9/093035} {\bibfield  {journal} {\bibinfo  {journal}
  {Nucl. Fusion}\ }\textbf {\bibinfo {volume} {55}},\ \bibinfo {pages} {093035}
  (\bibinfo {year} {2015})}\BibitemShut {NoStop}%
\bibitem [{\citenamefont {Pitts}\ \emph {et~al.}(2005)\citenamefont {Pitts},
  \citenamefont {Coad}, \citenamefont {Coster}, \citenamefont {Federici},
  \citenamefont {Fundamenski}, \citenamefont {Horacek}, \citenamefont
  {Krieger}, \citenamefont {Kukushkin}, \citenamefont {Likonen}, \citenamefont
  {Matthews}, \citenamefont {Rubel}, \citenamefont {Strachan},\ and\
  \citenamefont {Contributors}}]{pitts-2005}%
  \BibitemOpen
  \bibfield  {author} {\bibinfo {author} {\bibfnamefont {R.~A.}\ \bibnamefont
  {Pitts}}, \bibinfo {author} {\bibfnamefont {J.~P.}\ \bibnamefont {Coad}},
  \bibinfo {author} {\bibfnamefont {D.~P.}\ \bibnamefont {Coster}}, \bibinfo
  {author} {\bibfnamefont {G.}~\bibnamefont {Federici}}, \bibinfo {author}
  {\bibfnamefont {W.}~\bibnamefont {Fundamenski}}, \bibinfo {author}
  {\bibfnamefont {J.}~\bibnamefont {Horacek}}, \bibinfo {author} {\bibfnamefont
  {K.}~\bibnamefont {Krieger}}, \bibinfo {author} {\bibfnamefont
  {A.}~\bibnamefont {Kukushkin}}, \bibinfo {author} {\bibfnamefont
  {J.}~\bibnamefont {Likonen}}, \bibinfo {author} {\bibfnamefont {G.~F.}\
  \bibnamefont {Matthews}}, \bibinfo {author} {\bibfnamefont {M.}~\bibnamefont
  {Rubel}}, \bibinfo {author} {\bibfnamefont {J.~D.}\ \bibnamefont {Strachan}},
  \ and\ \bibinfo {author} {\bibfnamefont {J.-E.}\ \bibnamefont
  {Contributors}},\ }\href {\doibase 10.1088/0741-3335/47/12B/S22} {\bibfield
  {journal} {\bibinfo  {journal} {Plasma Phys. Control. Fusion}\ }\textbf
  {\bibinfo {volume} {47}},\ \bibinfo {pages} {B303} (\bibinfo {year}
  {2005})}\BibitemShut {NoStop}%
\bibitem [{\citenamefont {Lipschultz}\ \emph {et~al.}(2007)\citenamefont
  {Lipschultz}, \citenamefont {Bonnin}, \citenamefont {Counsell}, \citenamefont
  {Kallenbach}, \citenamefont {Kukushkin}, \citenamefont {Krieger},
  \citenamefont {Leonard}, \citenamefont {Loarte}, \citenamefont {Neu},
  \citenamefont {Pitts}, \citenamefont {Rognlien}, \citenamefont {Roth},
  \citenamefont {Skinner}, \citenamefont {Terry}, \citenamefont {Tsitrone},
  \citenamefont {Whyte}, \citenamefont {Zweben}, \citenamefont {Asakura},
  \citenamefont {Coster}, \citenamefont {Doerner}, \citenamefont {Dux},
  \citenamefont {Federici}, \citenamefont {Fenstermacher}, \citenamefont
  {Fundamenski}, \citenamefont {Ghendrih}, \citenamefont {Herrmann},
  \citenamefont {Hu}, \citenamefont {Krasheninnikov}, \citenamefont {Kirnev},
  \citenamefont {Kreter}, \citenamefont {Kurnaev}, \citenamefont {LaBombard},
  \citenamefont {Lisgo}, \citenamefont {Nakano}, \citenamefont {Ohno},
  \citenamefont {Pacher}, \citenamefont {Paley}, \citenamefont {Pan},
  \citenamefont {Pautasso}, \citenamefont {Philipps}, \citenamefont {Rohde},
  \citenamefont {Rudakov}, \citenamefont {Stangeby}, \citenamefont {Takamura},
  \citenamefont {Tanabe}, \citenamefont {Yang},\ and\ \citenamefont
  {Zhu}}]{lipschultz-2007}%
  \BibitemOpen
  \bibfield  {author} {\bibinfo {author} {\bibfnamefont {B.}~\bibnamefont
  {Lipschultz}}, \bibinfo {author} {\bibfnamefont {X.}~\bibnamefont {Bonnin}},
  \bibinfo {author} {\bibfnamefont {G.}~\bibnamefont {Counsell}}, \bibinfo
  {author} {\bibfnamefont {A.}~\bibnamefont {Kallenbach}}, \bibinfo {author}
  {\bibfnamefont {A.}~\bibnamefont {Kukushkin}}, \bibinfo {author}
  {\bibfnamefont {K.}~\bibnamefont {Krieger}}, \bibinfo {author} {\bibfnamefont
  {A.}~\bibnamefont {Leonard}}, \bibinfo {author} {\bibfnamefont
  {A.}~\bibnamefont {Loarte}}, \bibinfo {author} {\bibfnamefont
  {R.}~\bibnamefont {Neu}}, \bibinfo {author} {\bibfnamefont {R.}~\bibnamefont
  {Pitts}}, \bibinfo {author} {\bibfnamefont {T.}~\bibnamefont {Rognlien}},
  \bibinfo {author} {\bibfnamefont {J.}~\bibnamefont {Roth}}, \bibinfo {author}
  {\bibfnamefont {C.}~\bibnamefont {Skinner}}, \bibinfo {author} {\bibfnamefont
  {J.}~\bibnamefont {Terry}}, \bibinfo {author} {\bibfnamefont
  {E.}~\bibnamefont {Tsitrone}}, \bibinfo {author} {\bibfnamefont
  {D.}~\bibnamefont {Whyte}}, \bibinfo {author} {\bibfnamefont
  {S.}~\bibnamefont {Zweben}}, \bibinfo {author} {\bibfnamefont
  {N.}~\bibnamefont {Asakura}}, \bibinfo {author} {\bibfnamefont
  {D.}~\bibnamefont {Coster}}, \bibinfo {author} {\bibfnamefont
  {R.}~\bibnamefont {Doerner}}, \bibinfo {author} {\bibfnamefont
  {R.}~\bibnamefont {Dux}}, \bibinfo {author} {\bibfnamefont {G.}~\bibnamefont
  {Federici}}, \bibinfo {author} {\bibfnamefont {M.}~\bibnamefont
  {Fenstermacher}}, \bibinfo {author} {\bibfnamefont {W.}~\bibnamefont
  {Fundamenski}}, \bibinfo {author} {\bibfnamefont {P.}~\bibnamefont
  {Ghendrih}}, \bibinfo {author} {\bibfnamefont {A.}~\bibnamefont {Herrmann}},
  \bibinfo {author} {\bibfnamefont {J.}~\bibnamefont {Hu}}, \bibinfo {author}
  {\bibfnamefont {S.}~\bibnamefont {Krasheninnikov}}, \bibinfo {author}
  {\bibfnamefont {G.}~\bibnamefont {Kirnev}}, \bibinfo {author} {\bibfnamefont
  {A.}~\bibnamefont {Kreter}}, \bibinfo {author} {\bibfnamefont
  {V.}~\bibnamefont {Kurnaev}}, \bibinfo {author} {\bibfnamefont
  {B.}~\bibnamefont {LaBombard}}, \bibinfo {author} {\bibfnamefont
  {S.}~\bibnamefont {Lisgo}}, \bibinfo {author} {\bibfnamefont
  {T.}~\bibnamefont {Nakano}}, \bibinfo {author} {\bibfnamefont
  {N.}~\bibnamefont {Ohno}}, \bibinfo {author} {\bibfnamefont {H.}~\bibnamefont
  {Pacher}}, \bibinfo {author} {\bibfnamefont {J.}~\bibnamefont {Paley}},
  \bibinfo {author} {\bibfnamefont {Y.}~\bibnamefont {Pan}}, \bibinfo {author}
  {\bibfnamefont {G.}~\bibnamefont {Pautasso}}, \bibinfo {author}
  {\bibfnamefont {V.}~\bibnamefont {Philipps}}, \bibinfo {author}
  {\bibfnamefont {V.}~\bibnamefont {Rohde}}, \bibinfo {author} {\bibfnamefont
  {D.}~\bibnamefont {Rudakov}}, \bibinfo {author} {\bibfnamefont
  {P.}~\bibnamefont {Stangeby}}, \bibinfo {author} {\bibfnamefont
  {S.}~\bibnamefont {Takamura}}, \bibinfo {author} {\bibfnamefont
  {T.}~\bibnamefont {Tanabe}}, \bibinfo {author} {\bibfnamefont
  {Y.}~\bibnamefont {Yang}}, \ and\ \bibinfo {author} {\bibfnamefont
  {S.}~\bibnamefont {Zhu}},\ }\href {\doibase 10.1088/0029-5515/47/9/016}
  {\bibfield  {journal} {\bibinfo  {journal} {Nucl. Fusion}\ }\textbf {\bibinfo
  {volume} {47}},\ \bibinfo {pages} {1189} (\bibinfo {year}
  {2007})}\BibitemShut {NoStop}%
\bibitem [{\citenamefont {Garcia}\ \emph
  {et~al.}(2007{\natexlab{a}})\citenamefont {Garcia}, \citenamefont {Pitts},
  \citenamefont {Horacek}, \citenamefont {Nielsen}, \citenamefont
  {Fundamenski}, \citenamefont {Graves}, \citenamefont {Naulin},\ and\
  \citenamefont {Rasmussen}}]{garcia-jnm-2007}%
  \BibitemOpen
  \bibfield  {author} {\bibinfo {author} {\bibfnamefont {O.~E.}\ \bibnamefont
  {Garcia}}, \bibinfo {author} {\bibfnamefont {R.~A.}\ \bibnamefont {Pitts}},
  \bibinfo {author} {\bibfnamefont {J.}~\bibnamefont {Horacek}}, \bibinfo
  {author} {\bibfnamefont {A.~H.}\ \bibnamefont {Nielsen}}, \bibinfo {author}
  {\bibfnamefont {W.}~\bibnamefont {Fundamenski}}, \bibinfo {author}
  {\bibfnamefont {J.~P.}\ \bibnamefont {Graves}}, \bibinfo {author}
  {\bibfnamefont {V.}~\bibnamefont {Naulin}}, \ and\ \bibinfo {author}
  {\bibfnamefont {J.~J.}\ \bibnamefont {Rasmussen}},\ }\href {\doibase
  10.1016/j.jnucmat.2006.12.063} {\bibfield  {journal} {\bibinfo  {journal} {J.
  Nucl. Mater.}\ }\textbf {\bibinfo {volume} {363-365}},\ \bibinfo {pages}
  {575} (\bibinfo {year} {2007}{\natexlab{a}})}\BibitemShut {NoStop}%
\bibitem [{\citenamefont {Garcia}\ \emph
  {et~al.}(2007{\natexlab{b}})\citenamefont {Garcia}, \citenamefont {Horacek},
  \citenamefont {Pitts}, \citenamefont {Nielsen}, \citenamefont {Fundamenski},
  \citenamefont {Naulin},\ and\ \citenamefont {Rasmussen}}]{garcia-nf-2007}%
  \BibitemOpen
  \bibfield  {author} {\bibinfo {author} {\bibfnamefont {O.}~\bibnamefont
  {Garcia}}, \bibinfo {author} {\bibfnamefont {J.}~\bibnamefont {Horacek}},
  \bibinfo {author} {\bibfnamefont {R.}~\bibnamefont {Pitts}}, \bibinfo
  {author} {\bibfnamefont {A.}~\bibnamefont {Nielsen}}, \bibinfo {author}
  {\bibfnamefont {W.}~\bibnamefont {Fundamenski}}, \bibinfo {author}
  {\bibfnamefont {V.}~\bibnamefont {Naulin}}, \ and\ \bibinfo {author}
  {\bibfnamefont {J.~J.}\ \bibnamefont {Rasmussen}},\ }\href {\doibase
  10.1088/0029-5515/47/7/017} {\bibfield  {journal} {\bibinfo  {journal} {Nucl.
  Fusion}\ }\textbf {\bibinfo {volume} {47}},\ \bibinfo {pages} {667} (\bibinfo
  {year} {2007}{\natexlab{b}})}\BibitemShut {NoStop}%
\bibitem [{\citenamefont {Militello}\ \emph {et~al.}(2013)\citenamefont
  {Militello}, \citenamefont {Tamain}, \citenamefont {Fundamenski},
  \citenamefont {Kirk}, \citenamefont {Naulin},\ and\ \citenamefont
  {Nielsen}}]{militello-ppcf-2013}%
  \BibitemOpen
  \bibfield  {author} {\bibinfo {author} {\bibfnamefont {F.}~\bibnamefont
  {Militello}}, \bibinfo {author} {\bibfnamefont {P.}~\bibnamefont {Tamain}},
  \bibinfo {author} {\bibfnamefont {W.}~\bibnamefont {Fundamenski}}, \bibinfo
  {author} {\bibfnamefont {A.}~\bibnamefont {Kirk}}, \bibinfo {author}
  {\bibfnamefont {V.}~\bibnamefont {Naulin}}, \ and\ \bibinfo {author}
  {\bibfnamefont {A.~H.}\ \bibnamefont {Nielsen}},\ }\href {\doibase
  10.1088/0741-3335/55/2/025005} {\bibfield  {journal} {\bibinfo  {journal}
  {Plasma Phys. Control. Fusion}\ }\textbf {\bibinfo {volume} {55}},\ \bibinfo
  {pages} {025005} (\bibinfo {year} {2013})}\BibitemShut {NoStop}%
\bibitem [{\citenamefont {Carralero}\ \emph {et~al.}(2014)\citenamefont
  {Carralero}, \citenamefont {Birkenmeier}, \citenamefont {M{\"{u}}ller},
  \citenamefont {Manz}, \citenamefont {DeMarne}, \citenamefont {M{\"{u}}ller},
  \citenamefont {Reimold}, \citenamefont {Stroth}, \citenamefont {Wischmeier},\
  and\ \citenamefont {Wolfrum}}]{carralero-nf-2014}%
  \BibitemOpen
  \bibfield  {author} {\bibinfo {author} {\bibfnamefont {D.}~\bibnamefont
  {Carralero}}, \bibinfo {author} {\bibfnamefont {G.}~\bibnamefont
  {Birkenmeier}}, \bibinfo {author} {\bibfnamefont {H.}~\bibnamefont
  {M{\"{u}}ller}}, \bibinfo {author} {\bibfnamefont {P.}~\bibnamefont {Manz}},
  \bibinfo {author} {\bibfnamefont {P.}~\bibnamefont {DeMarne}}, \bibinfo
  {author} {\bibfnamefont {S.}~\bibnamefont {M{\"{u}}ller}}, \bibinfo {author}
  {\bibfnamefont {F.}~\bibnamefont {Reimold}}, \bibinfo {author} {\bibfnamefont
  {U.}~\bibnamefont {Stroth}}, \bibinfo {author} {\bibfnamefont
  {M.}~\bibnamefont {Wischmeier}}, \ and\ \bibinfo {author} {\bibfnamefont
  {E.}~\bibnamefont {Wolfrum}},\ }\href {\doibase
  10.1088/0029-5515/54/12/123005} {\bibfield  {journal} {\bibinfo  {journal}
  {Nucl. Fusion}\ }\textbf {\bibinfo {volume} {54}},\ \bibinfo {pages} {123005}
  (\bibinfo {year} {2014})}\BibitemShut {NoStop}%
\bibitem [{\citenamefont {Carralero}\ \emph {et~al.}(2015)\citenamefont
  {Carralero}, \citenamefont {Manz}, \citenamefont {Aho-Mantila}, \citenamefont
  {Birkenmeier}, \citenamefont {Brix}, \citenamefont {Groth}, \citenamefont
  {M{\"{u}}ller}, \citenamefont {Stroth}, \citenamefont {Vianello},\ and\
  \citenamefont {Wolfrum}}]{carralero-prl-2015}%
  \BibitemOpen
  \bibfield  {author} {\bibinfo {author} {\bibfnamefont {D.}~\bibnamefont
  {Carralero}}, \bibinfo {author} {\bibfnamefont {P.}~\bibnamefont {Manz}},
  \bibinfo {author} {\bibfnamefont {L.}~\bibnamefont {Aho-Mantila}}, \bibinfo
  {author} {\bibfnamefont {G.}~\bibnamefont {Birkenmeier}}, \bibinfo {author}
  {\bibfnamefont {M.}~\bibnamefont {Brix}}, \bibinfo {author} {\bibfnamefont
  {M.}~\bibnamefont {Groth}}, \bibinfo {author} {\bibfnamefont {H.~W.}\
  \bibnamefont {M{\"{u}}ller}}, \bibinfo {author} {\bibfnamefont
  {U.}~\bibnamefont {Stroth}}, \bibinfo {author} {\bibfnamefont
  {N.}~\bibnamefont {Vianello}}, \ and\ \bibinfo {author} {\bibfnamefont
  {E.}~\bibnamefont {Wolfrum}},\ }\href {\doibase
  10.1103/PhysRevLett.115.215002} {\bibfield  {journal} {\bibinfo  {journal}
  {Phys. Rev. Lett.}\ }\textbf {\bibinfo {volume} {115}},\ \bibinfo {pages}
  {215002} (\bibinfo {year} {2015})}\BibitemShut {NoStop}%
\bibitem [{\citenamefont {Militello}\ and\ \citenamefont
  {Omotani}(2016{\natexlab{a}})}]{militello-ppcf-2016}%
  \BibitemOpen
  \bibfield  {author} {\bibinfo {author} {\bibfnamefont {F.}~\bibnamefont
  {Militello}}\ and\ \bibinfo {author} {\bibfnamefont {J.~T.}\ \bibnamefont
  {Omotani}},\ }\href {\doibase 10.1088/0741-3335/58/12/125004} {\bibfield
  {journal} {\bibinfo  {journal} {Plasma Phys. Control. Fusion}\ }\textbf
  {\bibinfo {volume} {58}},\ \bibinfo {pages} {125004} (\bibinfo {year}
  {2016}{\natexlab{a}})}\BibitemShut {NoStop}%
\bibitem [{\citenamefont {Walkden}\ \emph
  {et~al.}(2017{\natexlab{a}})\citenamefont {Walkden}, \citenamefont {Wynn},
  \citenamefont {Militello}, \citenamefont {Lipschultz}, \citenamefont
  {Matthews}, \citenamefont {Guillemaut}, \citenamefont {Harrison},\ and\
  \citenamefont {Moulton}}]{walkden-ppcf-2017}%
  \BibitemOpen
  \bibfield  {author} {\bibinfo {author} {\bibfnamefont {N.~R.}\ \bibnamefont
  {Walkden}}, \bibinfo {author} {\bibfnamefont {A.}~\bibnamefont {Wynn}},
  \bibinfo {author} {\bibfnamefont {F.}~\bibnamefont {Militello}}, \bibinfo
  {author} {\bibfnamefont {B.}~\bibnamefont {Lipschultz}}, \bibinfo {author}
  {\bibfnamefont {G.}~\bibnamefont {Matthews}}, \bibinfo {author}
  {\bibfnamefont {C.}~\bibnamefont {Guillemaut}}, \bibinfo {author}
  {\bibfnamefont {J.}~\bibnamefont {Harrison}}, \ and\ \bibinfo {author}
  {\bibfnamefont {D.}~\bibnamefont {Moulton}},\ }\href {\doibase
  10.1088/1361-6587/aa7365} {\bibfield  {journal} {\bibinfo  {journal} {Plasma
  Phys. Control. Fusion}\ }\textbf {\bibinfo {volume} {59}},\ \bibinfo {pages}
  {085009} (\bibinfo {year} {2017}{\natexlab{a}})}\BibitemShut {NoStop}%
\bibitem [{\citenamefont {Garcia}\ \emph {et~al.}(2013)\citenamefont {Garcia},
  \citenamefont {Fritzner}, \citenamefont {Kube}, \citenamefont {Cziegler},
  \citenamefont {LaBombard},\ and\ \citenamefont {Terry}}]{garcia-pop-2013}%
  \BibitemOpen
  \bibfield  {author} {\bibinfo {author} {\bibfnamefont {O.~E.}\ \bibnamefont
  {Garcia}}, \bibinfo {author} {\bibfnamefont {S.~M.}\ \bibnamefont
  {Fritzner}}, \bibinfo {author} {\bibfnamefont {R.}~\bibnamefont {Kube}},
  \bibinfo {author} {\bibfnamefont {I.}~\bibnamefont {Cziegler}}, \bibinfo
  {author} {\bibfnamefont {B.}~\bibnamefont {LaBombard}}, \ and\ \bibinfo
  {author} {\bibfnamefont {J.~L.}\ \bibnamefont {Terry}},\ }\href {\doibase
  10.1063/1.4802942} {\bibfield  {journal} {\bibinfo  {journal} {Phys.
  Plasmas}\ }\textbf {\bibinfo {volume} {20}},\ \bibinfo {pages} {055901}
  (\bibinfo {year} {2013})}\BibitemShut {NoStop}%
\bibitem [{\citenamefont {Kube}\ \emph {et~al.}(2016)\citenamefont {Kube},
  \citenamefont {Theodorsen}, \citenamefont {Garcia}, \citenamefont
  {LaBombard},\ and\ \citenamefont {Terry}}]{kube-2016}%
  \BibitemOpen
  \bibfield  {author} {\bibinfo {author} {\bibfnamefont {R.}~\bibnamefont
  {Kube}}, \bibinfo {author} {\bibfnamefont {A.}~\bibnamefont {Theodorsen}},
  \bibinfo {author} {\bibfnamefont {O.~E.}\ \bibnamefont {Garcia}}, \bibinfo
  {author} {\bibfnamefont {B.}~\bibnamefont {LaBombard}}, \ and\ \bibinfo
  {author} {\bibfnamefont {J.~L.}\ \bibnamefont {Terry}},\ }\href {\doibase
  10.1088/0741-3335/58/5/054001} {\bibfield  {journal} {\bibinfo  {journal}
  {Plasma Phys. Control. Fusion}\ }\textbf {\bibinfo {volume} {58}},\ \bibinfo
  {pages} {054001} (\bibinfo {year} {2016})}\BibitemShut {NoStop}%
\bibitem [{\citenamefont {Garcia}\ \emph {et~al.}(2017)\citenamefont {Garcia},
  \citenamefont {Kube}, \citenamefont {Theodorsen}, \citenamefont {Bak},
  \citenamefont {Hong}, \citenamefont {Kim}, \citenamefont {Team},\ and\
  \citenamefont {Pitts}}]{garcia-nme-2016}%
  \BibitemOpen
  \bibfield  {author} {\bibinfo {author} {\bibfnamefont {O.}~\bibnamefont
  {Garcia}}, \bibinfo {author} {\bibfnamefont {R.}~\bibnamefont {Kube}},
  \bibinfo {author} {\bibfnamefont {A.}~\bibnamefont {Theodorsen}}, \bibinfo
  {author} {\bibfnamefont {J.-G.}\ \bibnamefont {Bak}}, \bibinfo {author}
  {\bibfnamefont {S.-H.}\ \bibnamefont {Hong}}, \bibinfo {author}
  {\bibfnamefont {H.-S.}\ \bibnamefont {Kim}}, \bibinfo {author} {\bibfnamefont
  {t.~K.~P.}\ \bibnamefont {Team}}, \ and\ \bibinfo {author} {\bibfnamefont
  {R.}~\bibnamefont {Pitts}},\ }\href {\doibase 10.1016/j.nme.2016.11.008}
  {\bibfield  {journal} {\bibinfo  {journal} {Nucl. Mater. Energy}\ }\textbf
  {\bibinfo {volume} {12}},\ \bibinfo {pages} {36} (\bibinfo {year}
  {2017})}\BibitemShut {NoStop}%
\bibitem [{\citenamefont {Theodorsen}\ \emph {et~al.}(2016)\citenamefont
  {Theodorsen}, \citenamefont {Garcia}, \citenamefont {Horacek}, \citenamefont
  {Kube},\ and\ \citenamefont {Pitts}}]{theodorsen-ppcf-2016}%
  \BibitemOpen
  \bibfield  {author} {\bibinfo {author} {\bibfnamefont {A.}~\bibnamefont
  {Theodorsen}}, \bibinfo {author} {\bibfnamefont {O.~E.}\ \bibnamefont
  {Garcia}}, \bibinfo {author} {\bibfnamefont {J.}~\bibnamefont {Horacek}},
  \bibinfo {author} {\bibfnamefont {R.}~\bibnamefont {Kube}}, \ and\ \bibinfo
  {author} {\bibfnamefont {R.~A.}\ \bibnamefont {Pitts}},\ }\href {\doibase
  10.1088/0741-3335/58/4/044006} {\bibfield  {journal} {\bibinfo  {journal}
  {Plasma Phys. Control. Fusion}\ }\textbf {\bibinfo {volume} {58}},\ \bibinfo
  {pages} {044006} (\bibinfo {year} {2016})}\BibitemShut {NoStop}%
\bibitem [{\citenamefont {Theodorsen}\ \emph
  {et~al.}(2017{\natexlab{a}})\citenamefont {Theodorsen}, \citenamefont
  {Garcia}, \citenamefont {Kube}, \citenamefont {LaBombard},\ and\
  \citenamefont {Terry}}]{theodorsen-nf-2017}%
  \BibitemOpen
  \bibfield  {author} {\bibinfo {author} {\bibfnamefont {A.}~\bibnamefont
  {Theodorsen}}, \bibinfo {author} {\bibfnamefont {O.}~\bibnamefont {Garcia}},
  \bibinfo {author} {\bibfnamefont {R.}~\bibnamefont {Kube}}, \bibinfo {author}
  {\bibfnamefont {B.}~\bibnamefont {LaBombard}}, \ and\ \bibinfo {author}
  {\bibfnamefont {J.}~\bibnamefont {Terry}},\ }\href {\doibase
  10.1088/1741-4326/aa7e4c} {\bibfield  {journal} {\bibinfo  {journal} {Nucl.
  Fusion}\ }\textbf {\bibinfo {volume} {57}},\ \bibinfo {pages} {114004}
  (\bibinfo {year} {2017}{\natexlab{a}})}\BibitemShut {NoStop}%
\bibitem [{\citenamefont {Kube}\ \emph {et~al.}()\citenamefont {Kube},
  \citenamefont {Garcia}, \citenamefont {Theodorsen}, \citenamefont {Brunner},
  \citenamefont {Kuang}, \citenamefont {LaBombard},\ and\ \citenamefont
  {Terry}}]{kube-cmod}%
  \BibitemOpen
  \bibfield  {author} {\bibinfo {author} {\bibfnamefont {R.}~\bibnamefont
  {Kube}}, \bibinfo {author} {\bibfnamefont {O.}~\bibnamefont {Garcia}},
  \bibinfo {author} {\bibfnamefont {A.}~\bibnamefont {Theodorsen}}, \bibinfo
  {author} {\bibfnamefont {D.}~\bibnamefont {Brunner}}, \bibinfo {author}
  {\bibfnamefont {A.}~\bibnamefont {Kuang}}, \bibinfo {author} {\bibfnamefont
  {B.}~\bibnamefont {LaBombard}}, \ and\ \bibinfo {author} {\bibfnamefont
  {J.}~\bibnamefont {Terry}},\ }\href@noop {} {\enquote {\bibinfo {title}
  {{Statistics of fluctuation driven transport in the scrape-off layer of
  Alcator C-Mod}},}\ }\bibinfo {note} {Unpublished}\BibitemShut {NoStop}%
\bibitem [{\citenamefont {Walkden}\ \emph
  {et~al.}(2017{\natexlab{b}})\citenamefont {Walkden}, \citenamefont {Wynn},
  \citenamefont {Militello}, \citenamefont {Lipschultz}, \citenamefont
  {Matthews}, \citenamefont {Guillemaut}, \citenamefont {Harrison},\ and\
  \citenamefont {Moulton}}]{walkden-nf-2017}%
  \BibitemOpen
  \bibfield  {author} {\bibinfo {author} {\bibfnamefont {N.}~\bibnamefont
  {Walkden}}, \bibinfo {author} {\bibfnamefont {A.}~\bibnamefont {Wynn}},
  \bibinfo {author} {\bibfnamefont {F.}~\bibnamefont {Militello}}, \bibinfo
  {author} {\bibfnamefont {B.}~\bibnamefont {Lipschultz}}, \bibinfo {author}
  {\bibfnamefont {G.}~\bibnamefont {Matthews}}, \bibinfo {author}
  {\bibfnamefont {C.}~\bibnamefont {Guillemaut}}, \bibinfo {author}
  {\bibfnamefont {J.}~\bibnamefont {Harrison}}, \ and\ \bibinfo {author}
  {\bibfnamefont {D.}~\bibnamefont {Moulton}},\ }\href {\doibase
  10.1088/1741-4326/aa515a} {\bibfield  {journal} {\bibinfo  {journal} {Nucl.
  Fusion}\ }\textbf {\bibinfo {volume} {57}},\ \bibinfo {pages} {036016}
  (\bibinfo {year} {2017}{\natexlab{b}})}\BibitemShut {NoStop}%
\bibitem [{\citenamefont {Kim}\ and\ \citenamefont
  {Anderson}(2008)}]{kim-2008}%
  \BibitemOpen
  \bibfield  {author} {\bibinfo {author} {\bibfnamefont {E.-j.}\ \bibnamefont
  {Kim}}\ and\ \bibinfo {author} {\bibfnamefont {J.}~\bibnamefont {Anderson}},\
  }\href {\doibase 10.1063/1.3033751} {\bibfield  {journal} {\bibinfo
  {journal} {Phys. Plasmas}\ }\textbf {\bibinfo {volume} {15}},\ \bibinfo
  {pages} {114506} (\bibinfo {year} {2008})}\BibitemShut {NoStop}%
\bibitem [{\citenamefont {Anderson}\ and\ \citenamefont
  {Xanthopoulos}(2010)}]{anderson-2010}%
  \BibitemOpen
  \bibfield  {author} {\bibinfo {author} {\bibfnamefont {J.}~\bibnamefont
  {Anderson}}\ and\ \bibinfo {author} {\bibfnamefont {P.}~\bibnamefont
  {Xanthopoulos}},\ }\href {\doibase 10.1063/1.3505824} {\bibfield  {journal}
  {\bibinfo  {journal} {Phys. Plasmas}\ }\textbf {\bibinfo {volume} {17}},\
  \bibinfo {pages} {110702} (\bibinfo {year} {2010})}\BibitemShut {NoStop}%
\bibitem [{\citenamefont {Rice}(1944)}]{rice-1944}%
  \BibitemOpen
  \bibfield  {author} {\bibinfo {author} {\bibfnamefont {S.~O.}\ \bibnamefont
  {Rice}},\ }\href {\doibase 10.1002/j.1538-7305.1944.tb00874.x} {\bibfield
  {journal} {\bibinfo  {journal} {Bell Syst. Tech. J.}\ }\textbf {\bibinfo
  {volume} {23}},\ \bibinfo {pages} {282} (\bibinfo {year} {1944})}\BibitemShut
  {NoStop}%
\bibitem [{\citenamefont {Rice}(1945)}]{rice-1945}%
  \BibitemOpen
  \bibfield  {author} {\bibinfo {author} {\bibfnamefont {S.~O.}\ \bibnamefont
  {Rice}},\ }\href {\doibase 10.1002/j.1538-7305.1945.tb00453.x} {\bibfield
  {journal} {\bibinfo  {journal} {Bell Syst. Tech. J.}\ }\textbf {\bibinfo
  {volume} {24}},\ \bibinfo {pages} {46} (\bibinfo {year} {1945})}\BibitemShut
  {NoStop}%
\bibitem [{\citenamefont {Fesce}(1986)}]{fesce-1986}%
  \BibitemOpen
  \bibfield  {author} {\bibinfo {author} {\bibfnamefont {R.}~\bibnamefont
  {Fesce}},\ }\href {\doibase 10.1085/jgp.88.1.25} {\bibfield  {journal}
  {\bibinfo  {journal} {J. Gen. Physiol.}\ }\textbf {\bibinfo {volume} {88}},\
  \bibinfo {pages} {25} (\bibinfo {year} {1986})}\BibitemShut {NoStop}%
\bibitem [{\citenamefont {Kristensen}\ \emph {et~al.}(1991)\citenamefont
  {Kristensen}, \citenamefont {Casanova}, \citenamefont {Courtney},\ and\
  \citenamefont {Troen}}]{kristensen-1991}%
  \BibitemOpen
  \bibfield  {author} {\bibinfo {author} {\bibfnamefont {L.}~\bibnamefont
  {Kristensen}}, \bibinfo {author} {\bibfnamefont {M.}~\bibnamefont
  {Casanova}}, \bibinfo {author} {\bibfnamefont {M.~S.}\ \bibnamefont
  {Courtney}}, \ and\ \bibinfo {author} {\bibfnamefont {I.}~\bibnamefont
  {Troen}},\ }\href {\doibase 10.1007/BF00119328} {\bibfield  {journal}
  {\bibinfo  {journal} {Boundary-Layer Meteorol.}\ }\textbf {\bibinfo {volume}
  {55}},\ \bibinfo {pages} {91} (\bibinfo {year} {1991})}\BibitemShut {NoStop}%
\bibitem [{\citenamefont {Jang}(2004)}]{jang-2004}%
  \BibitemOpen
  \bibfield  {author} {\bibinfo {author} {\bibfnamefont {J.-W.}\ \bibnamefont
  {Jang}},\ }\href {\doibase 10.1111/j.0022-4367.2004.00086.x} {\bibfield
  {journal} {\bibinfo  {journal} {J. Risk Insur.}\ }\textbf {\bibinfo {volume}
  {71}},\ \bibinfo {pages} {201} (\bibinfo {year} {2004})}\BibitemShut
  {NoStop}%
\bibitem [{\citenamefont {Daly}\ and\ \citenamefont
  {Porporato}(2006)}]{daly-2006}%
  \BibitemOpen
  \bibfield  {author} {\bibinfo {author} {\bibfnamefont {E.}~\bibnamefont
  {Daly}}\ and\ \bibinfo {author} {\bibfnamefont {A.}~\bibnamefont
  {Porporato}},\ }\href {\doibase 10.1103/PhysRevE.73.026108} {\bibfield
  {journal} {\bibinfo  {journal} {Phys. Rev. E}\ }\textbf {\bibinfo {volume}
  {73}},\ \bibinfo {pages} {026108} (\bibinfo {year} {2006})}\BibitemShut
  {NoStop}%
\bibitem [{\citenamefont {Narasimha}\ \emph {et~al.}(2007)\citenamefont
  {Narasimha}, \citenamefont {Kumar}, \citenamefont {Prabhu},\ and\
  \citenamefont {Kailas}}]{narasimha-2007}%
  \BibitemOpen
  \bibfield  {author} {\bibinfo {author} {\bibfnamefont {R.}~\bibnamefont
  {Narasimha}}, \bibinfo {author} {\bibfnamefont {S.~R.}\ \bibnamefont
  {Kumar}}, \bibinfo {author} {\bibfnamefont {A.}~\bibnamefont {Prabhu}}, \
  and\ \bibinfo {author} {\bibfnamefont {S.~V.}\ \bibnamefont {Kailas}},\
  }\href {\doibase 10.1098/rsta.2006.1949} {\bibfield  {journal} {\bibinfo
  {journal} {Philos. Trans. A. Math. Phys. Eng. Sci.}\ }\textbf {\bibinfo
  {volume} {365}},\ \bibinfo {pages} {841} (\bibinfo {year}
  {2007})}\BibitemShut {NoStop}%
\bibitem [{\citenamefont {Elter}\ \emph {et~al.}(2015)\citenamefont {Elter},
  \citenamefont {Jammes}, \citenamefont {P{\'{a}}zsit}, \citenamefont
  {P{\'{a}}l},\ and\ \citenamefont {Filliatre}}]{elter-2015}%
  \BibitemOpen
  \bibfield  {author} {\bibinfo {author} {\bibfnamefont {Z.}~\bibnamefont
  {Elter}}, \bibinfo {author} {\bibfnamefont {C.}~\bibnamefont {Jammes}},
  \bibinfo {author} {\bibfnamefont {I.}~\bibnamefont {P{\'{a}}zsit}}, \bibinfo
  {author} {\bibfnamefont {L.}~\bibnamefont {P{\'{a}}l}}, \ and\ \bibinfo
  {author} {\bibfnamefont {P.}~\bibnamefont {Filliatre}},\ }\href {\doibase
  10.1016/j.nima.2014.11.065} {\bibfield  {journal} {\bibinfo  {journal} {Nucl.
  Instruments Methods Phys. Res. Sect. A Accel. Spectrometers, Detect. Assoc.
  Equip.}\ }\textbf {\bibinfo {volume} {774}},\ \bibinfo {pages} {60} (\bibinfo
  {year} {2015})}\BibitemShut {NoStop}%
\bibitem [{\citenamefont {Garcia}(2012)}]{garcia-prl-2012}%
  \BibitemOpen
  \bibfield  {author} {\bibinfo {author} {\bibfnamefont {O.~E.}\ \bibnamefont
  {Garcia}},\ }\href {\doibase 10.1103/PhysRevLett.108.265001} {\bibfield
  {journal} {\bibinfo  {journal} {Phys. Rev. Lett.}\ }\textbf {\bibinfo
  {volume} {108}},\ \bibinfo {pages} {265001} (\bibinfo {year}
  {2012})}\BibitemShut {NoStop}%
\bibitem [{\citenamefont {Garcia}\ \emph {et~al.}(2016)\citenamefont {Garcia},
  \citenamefont {Kube}, \citenamefont {Theodorsen},\ and\ \citenamefont
  {P{\'{e}}cseli}}]{garcia-pop-2016}%
  \BibitemOpen
  \bibfield  {author} {\bibinfo {author} {\bibfnamefont {O.~E.}\ \bibnamefont
  {Garcia}}, \bibinfo {author} {\bibfnamefont {R.}~\bibnamefont {Kube}},
  \bibinfo {author} {\bibfnamefont {A.}~\bibnamefont {Theodorsen}}, \ and\
  \bibinfo {author} {\bibfnamefont {H.~L.}\ \bibnamefont {P{\'{e}}cseli}},\
  }\href {\doibase 10.1063/1.4951016} {\bibfield  {journal} {\bibinfo
  {journal} {Phys. Plasmas}\ }\textbf {\bibinfo {volume} {23}},\ \bibinfo
  {pages} {052308} (\bibinfo {year} {2016})}\BibitemShut {NoStop}%
\bibitem [{\citenamefont {Militello}\ and\ \citenamefont
  {Omotani}(2016{\natexlab{b}})}]{militello-nf-2016}%
  \BibitemOpen
  \bibfield  {author} {\bibinfo {author} {\bibfnamefont {F.}~\bibnamefont
  {Militello}}\ and\ \bibinfo {author} {\bibfnamefont {J.}~\bibnamefont
  {Omotani}},\ }\href {\doibase 10.1088/0029-5515/56/10/104004} {\bibfield
  {journal} {\bibinfo  {journal} {Nucl. Fusion}\ }\textbf {\bibinfo {volume}
  {56}},\ \bibinfo {pages} {104004} (\bibinfo {year}
  {2016}{\natexlab{b}})}\BibitemShut {NoStop}%
\bibitem [{\citenamefont {Kube}\ and\ \citenamefont
  {Garcia}(2015)}]{kube-2015}%
  \BibitemOpen
  \bibfield  {author} {\bibinfo {author} {\bibfnamefont {R.}~\bibnamefont
  {Kube}}\ and\ \bibinfo {author} {\bibfnamefont {O.~E.}\ \bibnamefont
  {Garcia}},\ }\href {\doibase 10.1063/1.4905513} {\bibfield  {journal}
  {\bibinfo  {journal} {Phys. Plasmas}\ }\textbf {\bibinfo {volume} {22}},\
  \bibinfo {pages} {012502} (\bibinfo {year} {2015})}\BibitemShut {NoStop}%
\bibitem [{\citenamefont {Theodorsen}\ \emph
  {et~al.}(2017{\natexlab{b}})\citenamefont {Theodorsen}, \citenamefont
  {Garcia},\ and\ \citenamefont {Rypdal}}]{theodorsen-ps-2017}%
  \BibitemOpen
  \bibfield  {author} {\bibinfo {author} {\bibfnamefont {A.}~\bibnamefont
  {Theodorsen}}, \bibinfo {author} {\bibfnamefont {O.~E.}\ \bibnamefont
  {Garcia}}, \ and\ \bibinfo {author} {\bibfnamefont {M.}~\bibnamefont
  {Rypdal}},\ }\href {\doibase 10.1088/1402-4896/aa694c} {\bibfield  {journal}
  {\bibinfo  {journal} {Phys. Scr.}\ }\textbf {\bibinfo {volume} {92}},\
  \bibinfo {pages} {054002} (\bibinfo {year} {2017}{\natexlab{b}})}\BibitemShut
  {NoStop}%
\bibitem [{\citenamefont {Garcia}\ and\ \citenamefont
  {Theodorsen}(2017{\natexlab{a}})}]{garcia-pop-2017-1}%
  \BibitemOpen
  \bibfield  {author} {\bibinfo {author} {\bibfnamefont {O.~E.}\ \bibnamefont
  {Garcia}}\ and\ \bibinfo {author} {\bibfnamefont {A.}~\bibnamefont
  {Theodorsen}},\ }\href {\doibase 10.1063/1.4975645} {\bibfield  {journal}
  {\bibinfo  {journal} {Phys. Plasmas}\ }\textbf {\bibinfo {volume} {24}},\
  \bibinfo {pages} {020704} (\bibinfo {year} {2017}{\natexlab{a}})}\BibitemShut
  {NoStop}%
\bibitem [{\citenamefont {Garcia}\ and\ \citenamefont
  {Theodorsen}(2017{\natexlab{b}})}]{garcia-pop-2017-2}%
  \BibitemOpen
  \bibfield  {author} {\bibinfo {author} {\bibfnamefont {O.~E.}\ \bibnamefont
  {Garcia}}\ and\ \bibinfo {author} {\bibfnamefont {A.}~\bibnamefont
  {Theodorsen}},\ }\href {\doibase 10.1063/1.4978955} {\bibfield  {journal}
  {\bibinfo  {journal} {Phys. Plasmas}\ }\textbf {\bibinfo {volume} {24}},\
  \bibinfo {pages} {032309} (\bibinfo {year} {2017}{\natexlab{b}})}\BibitemShut
  {NoStop}%
\bibitem [{\citenamefont {Theodorsen}\ and\ \citenamefont
  {Garcia}(2016)}]{theodorsen-pop-2016}%
  \BibitemOpen
  \bibfield  {author} {\bibinfo {author} {\bibfnamefont {A.}~\bibnamefont
  {Theodorsen}}\ and\ \bibinfo {author} {\bibfnamefont {O.~E.}\ \bibnamefont
  {Garcia}},\ }\href {\doibase 10.1063/1.4947235} {\bibfield  {journal}
  {\bibinfo  {journal} {Phys. Plasmas}\ }\textbf {\bibinfo {volume} {23}},\
  \bibinfo {pages} {040702} (\bibinfo {year} {2016})}\BibitemShut {NoStop}%
\bibitem [{\citenamefont {Parzen}(1999)}]{parzen-sp}%
  \BibitemOpen
  \bibfield  {author} {\bibinfo {author} {\bibfnamefont {E.}~\bibnamefont
  {Parzen}},\ }\href {\doibase 10.1137/1.9781611971125} {\emph {\bibinfo
  {title} {{Stochastic Processes}}}}\ (\bibinfo  {publisher} {Society for
  Industrial and Applied Mathematics},\ \bibinfo {year} {1999})\BibitemShut
  {NoStop}%
\bibitem [{\citenamefont {P{\'{e}}cseli}(2000)}]{pecseli-fps}%
  \BibitemOpen
  \bibfield  {author} {\bibinfo {author} {\bibfnamefont {H.}~\bibnamefont
  {P{\'{e}}cseli}},\ }\href@noop {} {\emph {\bibinfo {title} {{Fluctuations in
  physical systems}}}}\ (\bibinfo  {publisher} {Cambridge University Press},\
  \bibinfo {year} {2000})\BibitemShut {NoStop}%
\bibitem [{\citenamefont {Sattin}\ \emph {et~al.}(2009)\citenamefont {Sattin},
  \citenamefont {Agostini}, \citenamefont {Cavazzana}, \citenamefont
  {Serianni}, \citenamefont {Scarin},\ and\ \citenamefont
  {Vianello}}]{sattin-2009}%
  \BibitemOpen
  \bibfield  {author} {\bibinfo {author} {\bibfnamefont {F.}~\bibnamefont
  {Sattin}}, \bibinfo {author} {\bibfnamefont {M.}~\bibnamefont {Agostini}},
  \bibinfo {author} {\bibfnamefont {R.}~\bibnamefont {Cavazzana}}, \bibinfo
  {author} {\bibfnamefont {G.}~\bibnamefont {Serianni}}, \bibinfo {author}
  {\bibfnamefont {P.}~\bibnamefont {Scarin}}, \ and\ \bibinfo {author}
  {\bibfnamefont {N.}~\bibnamefont {Vianello}},\ }\href {\doibase
  10.1088/0031-8949/79/04/045006} {\bibfield  {journal} {\bibinfo  {journal}
  {Phys. Scr.}\ }\textbf {\bibinfo {volume} {79}},\ \bibinfo {pages} {045006}
  (\bibinfo {year} {2009})}\BibitemShut {NoStop}%
\bibitem [{\citenamefont {Bergsaker}\ \emph {et~al.}(2015)\citenamefont
  {Bergsaker}, \citenamefont {Fredriksen}, \citenamefont {P{\'{e}}cseli},\ and\
  \citenamefont {Trulsen}}]{bergsaker-2015}%
  \BibitemOpen
  \bibfield  {author} {\bibinfo {author} {\bibfnamefont {A.~S.}\ \bibnamefont
  {Bergsaker}}, \bibinfo {author} {\bibfnamefont {{\AA}.}~\bibnamefont
  {Fredriksen}}, \bibinfo {author} {\bibfnamefont {H.~L.}\ \bibnamefont
  {P{\'{e}}cseli}}, \ and\ \bibinfo {author} {\bibfnamefont {J.~K.}\
  \bibnamefont {Trulsen}},\ }\href {\doibase 10.1088/0031-8949/90/10/108005}
  {\bibfield  {journal} {\bibinfo  {journal} {Phys. Scr.}\ }\textbf {\bibinfo
  {volume} {90}},\ \bibinfo {pages} {108005} (\bibinfo {year}
  {2015})}\BibitemShut {NoStop}%
\bibitem [{\citenamefont {Kozubowski}\ and\ \citenamefont
  {Podg{\'{o}}rski}(2000)}]{kozubowski-2000}%
  \BibitemOpen
  \bibfield  {author} {\bibinfo {author} {\bibfnamefont {T.~J.}\ \bibnamefont
  {Kozubowski}}\ and\ \bibinfo {author} {\bibfnamefont {K.}~\bibnamefont
  {Podg{\'{o}}rski}},\ }\href {\doibase 10.1007/PL00022717} {\bibfield
  {journal} {\bibinfo  {journal} {Comput. Stat.}\ }\textbf {\bibinfo {volume}
  {15}},\ \bibinfo {pages} {531} (\bibinfo {year} {2000})}\BibitemShut
  {NoStop}%
\bibitem [{\citenamefont {Reed}(2006)}]{reed-2006}%
  \BibitemOpen
  \bibfield  {author} {\bibinfo {author} {\bibfnamefont {W.~J.}\ \bibnamefont
  {Reed}},\ }in\ \href {\doibase 10.1007/0-8176-4487-3_4} {\emph {\bibinfo
  {booktitle} {Adv. Distrib. Theory, Order Stat. Inference}}}\ (\bibinfo
  {publisher} {Birkh{\"{a}}user Boston},\ \bibinfo {address} {Boston, MA},\
  \bibinfo {year} {2006})\ pp.\ \bibinfo {pages} {61--74}\BibitemShut {NoStop}%
\bibitem [{\citenamefont {Bondesson}(1982)}]{bondesson-1982}%
  \BibitemOpen
  \bibfield  {author} {\bibinfo {author} {\bibfnamefont {L.}~\bibnamefont
  {Bondesson}},\ }\href
  {http://www.jstor.org/stable/1427027{\%}5Cnpapers2://publication/uuid/73A926C7-A3AB-4CB8-88E9-F877E9257419}
  {\bibfield  {journal} {\bibinfo  {journal} {Adv. Appl. Probab.}\ }\textbf
  {\bibinfo {volume} {14}},\ \bibinfo {pages} {855} (\bibinfo {year}
  {1982})}\BibitemShut {NoStop}%
\bibitem [{\citenamefont {Daly}\ and\ \citenamefont
  {Porporato}(2010)}]{daly-2010}%
  \BibitemOpen
  \bibfield  {author} {\bibinfo {author} {\bibfnamefont {E.}~\bibnamefont
  {Daly}}\ and\ \bibinfo {author} {\bibfnamefont {A.}~\bibnamefont
  {Porporato}},\ }\href {\doibase 10.1103/PhysRevE.81.061133} {\bibfield
  {journal} {\bibinfo  {journal} {Phys. Rev. E}\ }\textbf {\bibinfo {volume}
  {81}},\ \bibinfo {pages} {061133} (\bibinfo {year} {2010})}\BibitemShut
  {NoStop}%
\bibitem [{\citenamefont {Olver}\ \emph {et~al.}()\citenamefont {Olver},
  \citenamefont {{Olde Daalhuis}}, \citenamefont {Lozier}, \citenamefont
  {Schneider}, \citenamefont {Boisvert}, \citenamefont {Clark}, \citenamefont
  {Miller},\ and\ \citenamefont {Saunders}}]{nist-dlmf}%
  \BibitemOpen
  \bibfield  {author} {\bibinfo {author} {\bibfnamefont {F.~W.~J.}\
  \bibnamefont {Olver}}, \bibinfo {author} {\bibfnamefont {A.~B.}\ \bibnamefont
  {{Olde Daalhuis}}}, \bibinfo {author} {\bibfnamefont {D.~W.}\ \bibnamefont
  {Lozier}}, \bibinfo {author} {\bibfnamefont {B.~I.}\ \bibnamefont
  {Schneider}}, \bibinfo {author} {\bibfnamefont {R.~F.}\ \bibnamefont
  {Boisvert}}, \bibinfo {author} {\bibfnamefont {C.~W.}\ \bibnamefont {Clark}},
  \bibinfo {author} {\bibfnamefont {B.~R.}\ \bibnamefont {Miller}}, \ and\
  \bibinfo {author} {\bibfnamefont {B.~V.}\ \bibnamefont {Saunders}},\ }\href
  {http://dlmf.nist.gov/} {\enquote {\bibinfo {title} {{NIST Digital Library of
  Mathematical Functions}},}\ }\bibinfo {howpublished} {http://dlmf.nist.gov/,
  Release 1.0.14 of 2016-12-21}\BibitemShut {NoStop}%
\bibitem [{\citenamefont {Feuerverger}\ and\ \citenamefont
  {Mureika}(1977)}]{feuerverger-1977}%
  \BibitemOpen
  \bibfield  {author} {\bibinfo {author} {\bibfnamefont {A.}~\bibnamefont
  {Feuerverger}}\ and\ \bibinfo {author} {\bibfnamefont {R.~A.}\ \bibnamefont
  {Mureika}},\ }\href {\doibase 10.1214/aos/1176343742} {\bibfield  {journal}
  {\bibinfo  {journal} {Ann. Stat.}\ }\textbf {\bibinfo {volume} {5}},\
  \bibinfo {pages} {88} (\bibinfo {year} {1977})}\BibitemShut {NoStop}%
\bibitem [{\citenamefont {Feuerverger}\ and\ \citenamefont
  {McDunnough}(1981)}]{feuerverger-1981}%
  \BibitemOpen
  \bibfield  {author} {\bibinfo {author} {\bibfnamefont {A.}~\bibnamefont
  {Feuerverger}}\ and\ \bibinfo {author} {\bibfnamefont {P.}~\bibnamefont
  {McDunnough}},\ }\href {http://www.jstor.org/stable/2985144} {\bibfield
  {journal} {\bibinfo  {journal} {J. R. Stat. Soc. Ser. B}\ }\textbf {\bibinfo
  {volume} {43}},\ \bibinfo {pages} {20} (\bibinfo {year} {1981})}\BibitemShut
  {NoStop}%
\bibitem [{\citenamefont {Yu}(2004)}]{yu-2004}%
  \BibitemOpen
  \bibfield  {author} {\bibinfo {author} {\bibfnamefont {J.}~\bibnamefont
  {Yu}},\ }\href {\doibase 10.1081/ETC-120039605} {\bibfield  {journal}
  {\bibinfo  {journal} {Econom. Rev.}\ }\textbf {\bibinfo {volume} {23}},\
  \bibinfo {pages} {93} (\bibinfo {year} {2004})}\BibitemShut {NoStop}%
\bibitem [{\citenamefont {Tran}(1998)}]{tran-1998}%
  \BibitemOpen
  \bibfield  {author} {\bibinfo {author} {\bibfnamefont {K.~C.}\ \bibnamefont
  {Tran}},\ }\href {\doibase 10.1080/07474939808800410} {\bibfield  {journal}
  {\bibinfo  {journal} {Econom. Rev.}\ }\textbf {\bibinfo {volume} {17}},\
  \bibinfo {pages} {167} (\bibinfo {year} {1998})}\BibitemShut {NoStop}%
\bibitem [{\citenamefont {Zhang}\ and\ \citenamefont {He}(2016)}]{zhang-2016}%
  \BibitemOpen
  \bibfield  {author} {\bibinfo {author} {\bibfnamefont {S.}~\bibnamefont
  {Zhang}}\ and\ \bibinfo {author} {\bibfnamefont {X.}~\bibnamefont {He}},\
  }\href {\doibase 10.1080/02331888.2015.1085866} {\bibfield  {journal}
  {\bibinfo  {journal} {Statistics (Ber).}\ }\textbf {\bibinfo {volume} {50}},\
  \bibinfo {pages} {667} (\bibinfo {year} {2016})}\BibitemShut {NoStop}%
\bibitem [{\citenamefont {Byrd}\ \emph {et~al.}(1995)\citenamefont {Byrd},
  \citenamefont {Lu}, \citenamefont {Nocedal},\ and\ \citenamefont
  {Zhu}}]{byrd-1995}%
  \BibitemOpen
  \bibfield  {author} {\bibinfo {author} {\bibfnamefont {R.~H.}\ \bibnamefont
  {Byrd}}, \bibinfo {author} {\bibfnamefont {P.}~\bibnamefont {Lu}}, \bibinfo
  {author} {\bibfnamefont {J.}~\bibnamefont {Nocedal}}, \ and\ \bibinfo
  {author} {\bibfnamefont {C.}~\bibnamefont {Zhu}},\ }\href {\doibase
  10.1137/0916069} {\bibfield  {journal} {\bibinfo  {journal} {SIAM J. Sci.
  Comput.}\ }\textbf {\bibinfo {volume} {16}},\ \bibinfo {pages} {1190}
  (\bibinfo {year} {1995})}\BibitemShut {NoStop}%
\bibitem [{\citenamefont {Zhu}\ \emph {et~al.}(1997)\citenamefont {Zhu},
  \citenamefont {Byrd}, \citenamefont {Lu},\ and\ \citenamefont
  {Nocedal}}]{zhu-1997}%
  \BibitemOpen
  \bibfield  {author} {\bibinfo {author} {\bibfnamefont {C.}~\bibnamefont
  {Zhu}}, \bibinfo {author} {\bibfnamefont {R.~H.}\ \bibnamefont {Byrd}},
  \bibinfo {author} {\bibfnamefont {P.}~\bibnamefont {Lu}}, \ and\ \bibinfo
  {author} {\bibfnamefont {J.}~\bibnamefont {Nocedal}},\ }\href {\doibase
  10.1145/279232.279236} {\bibfield  {journal} {\bibinfo  {journal} {ACM Trans.
  Math. Softw.}\ }\textbf {\bibinfo {volume} {23}},\ \bibinfo {pages} {550}
  (\bibinfo {year} {1997})}\BibitemShut {NoStop}%
\bibitem [{\citenamefont {Jones}\ \emph {et~al.}()\citenamefont {Jones},
  \citenamefont {Oliphant},\ and\ \citenamefont {Peterson}}]{scipy}%
  \BibitemOpen
  \bibfield  {author} {\bibinfo {author} {\bibfnamefont {E.}~\bibnamefont
  {Jones}}, \bibinfo {author} {\bibfnamefont {T.}~\bibnamefont {Oliphant}}, \
  and\ \bibinfo {author} {\bibfnamefont {P.}~\bibnamefont {Peterson}},\ }\href
  {http://www.scipy.org} {\enquote {\bibinfo {title} {{{SciPy}: Open source
  scientific tools for {Python}}},}\ }\bibinfo {howpublished}
  {http://www.scipy.org, Release 0.19.1 of 2017-06-21}\BibitemShut {NoStop}%
\bibitem [{\citenamefont {Lowen}\ and\ \citenamefont
  {Teich}(2005)}]{lowen-teich-fractal}%
  \BibitemOpen
  \bibfield  {author} {\bibinfo {author} {\bibfnamefont {S.~B.}\ \bibnamefont
  {Lowen}}\ and\ \bibinfo {author} {\bibfnamefont {M.~C.}\ \bibnamefont
  {Teich}},\ }\href {\doibase 10.1002/0471754722} {\emph {\bibinfo {title}
  {{Fractal-Based Point Processes}}}},\ Wiley Series in Probability and
  Statistics\ (\bibinfo  {publisher} {John Wiley \& Sons, Inc.},\ \bibinfo
  {address} {Hoboken, NJ, USA},\ \bibinfo {year} {2005})\BibitemShut {NoStop}%
\bibitem [{\citenamefont {{Stark, H.: Woods}}(2012)}]{stark-psrpe}%
  \BibitemOpen
  \bibfield  {author} {\bibinfo {author} {\bibfnamefont {J.~W.}\ \bibnamefont
  {{Stark, H.: Woods}}},\ }\href@noop {} {\emph {\bibinfo {title}
  {{Probability, Statistics and Random Processes for Engineers}}}},\ \bibinfo
  {edition} {4th}\ ed.\ (\bibinfo  {publisher} {Pearson Education},\ \bibinfo
  {year} {2012})\BibitemShut {NoStop}%
\end{thebibliography}
\end{document}